\documentclass[parskip=half]{scrartcl}
\title{Exact Bayesian Inference for Markov Switching Diffusions}
\author{
  Timothée Stumpf-Fétizon \\ \href{mailto:timothee.stumpffetizon@unibocconi.it}{\texttt{[timothee.stumpffetizon@unibocconi.it]}} \and
  Krzysztof Łatuszyński \\
  \href{mailto:k.g.latuszynski@warwick.ac.uk}{\texttt{[k.g.latuszynski@warwick.ac.uk]}} \and
  Jan Palczewski \\
  \href{mailto:j.palczewski@leeds.ac.uk}{\texttt{[j.palczewski@leeds.ac.uk]}} \and
  Gareth Roberts \\
  \href{mailto:gareth.o.roberts@warwick.ac.uk}{\texttt{[gareth.o.roberts@warwick.ac.uk]}}
}
    
\usepackage[noend]{algpseudocode}
\usepackage{algorithm}
\usepackage{amssymb}
\usepackage{amsthm}
\usepackage[backend=biber]{biblatex}
\usepackage{booktabs}
\usepackage[colorlinks, citecolor=blue, urlcolor=blue]{hyperref}
\usepackage{listings}
\usepackage{mathtools}
\usepackage{oubraces}
\usepackage{subcaption}
\usepackage{tikz}
\usepackage{todonotes}
\usepackage{xparse}

\usepackage{accents}

\NewDocumentCommand\ub{m}{{#1}^{\uparrow}}
\NewDocumentCommand\lb{m}{{#1}^{\downarrow}}

\NewDocumentCommand\est{m}{\bar{#1}}

\NewDocumentCommand\class{m}{\mathcal{#1}}
\NewDocumentCommand\meas{m}{\mathbb{#1}}

\NewDocumentCommand\leb{}{\mathrm{Leb}}

\DeclarePairedDelimiter{\braces}{\lbrace}{\rbrace}
\DeclarePairedDelimiter{\parens}{\lparen}{\rparen}

\DeclarePairedDelimiter{\angls}{\langle}{\rangle}
\DeclarePairedDelimiter{\bars}{\lvert}{\rvert}
\DeclarePairedDelimiter\ceil{\lceil}{\rceil}
\DeclarePairedDelimiter\floor{\lfloor}{\rfloor}

\NewDocumentCommand\setdiff{}{\mathbin{\backslash}}
\NewDocumentCommand\dd{g}{\mathop{}\!\mathrm{d} #1}
\NewDocumentCommand\der{gm}{\frac{\dd{} #1}{\dd{} #2}}
\NewDocumentCommand\pder{gm}{\frac{\partial #1}{\partial #2}}
\NewDocumentCommand\op{smO{}O{}g}{
  \IfBooleanTF{#1}{
    \IfNoValueTF{#5}
      {\operatorname*{\mathrm{#2}}_{#3}^{#4}}
      {\operatorname*{\mathrm{#2}}_{#3}^{#4} \left[ #5 \right]}
    }{
    \IfNoValueTF{#5}
      {\operatorname{\mathrm{#2}}_{#3}^{#4}}
      {\operatorname{\mathrm{#2}}_{#3}^{#4} \left[ #5 \right]}
    }
}

\NewDocumentCommand\reals{}{\mathbf{R}}

\theoremstyle{plain}

\newtheorem{proposition}{Proposition}

\definecolor{lstbase}{HTML}{555753}
\definecolor{lstkey}{HTML}{5C3566}
\definecolor{lstbg}{HTML}{EEEEEC}
\definecolor{lstcom}{HTML}{888A85}
\definecolor{lstnum}{HTML}{BABDB6}
\definecolor{lstid}{HTML}{555753}
\definecolor{lststr}{HTML}{8F5902}

\makeatletter
\newenvironment{breakablealgorithm}
  {
   \begin{center}
     \refstepcounter{algorithm}
     \renewcommand{\caption}[2][\relax]{
       {\raggedright\textbf{\ALG@name~\thealgorithm} ##2\par}%
       \ifx\relax##1\relax 
         \addcontentsline{loa}{algorithm}{\protect\numberline{\thealgorithm}##2}%
       \else 
         \addcontentsline{loa}{algorithm}{\protect\numberline{\thealgorithm}##1}%
       \fi
       \kern2pt\hrule\kern2pt
     }
  }{
     \kern2pt\hrule\relax
   \end{center}
  }
\makeatother
\let\citep = \cite

\begin{document}
\maketitle

\begin{abstract}
  We develop the first exact Bayesian methodology for the problem of inference in discretely observed regime switching diffusions. Switching diffusion models extend ordinary diffusions by allowing for jumps in instantaneous drift and volatility. The jumps are driven by a latent, continuous time Markov switching process. We address the problem through an MCMC and an MCEM algorithm that target the exact posterior of diffusion parameters and the latent regime process. The algorithms are exact in the sense that they target the correct posterior distribution of the continuous model, so that the errors are due to Monte Carlo only. We illustrate the method on numerical examples, including an empirical analysis of the method's scalability in the length of the time series, and find that it is comparable in computational cost with discrete approximations while avoiding their shortcomings.
  \par\vskip\baselineskip\noindent
  \textbf{Code:} \url{https://github.com/timsf/markov-switch}
  \par\vskip\baselineskip\noindent
  \textbf{Keywords:} diffusions, time series, regime switching, intractable likelihood, Bernoulli MCMC, Markov chain Monte Carlo.
  \par\vskip\baselineskip\noindent
  \textbf{MSC2020 subject classifications:} Primary 62-08; Secondary 62F15, 62M05
\end{abstract}

\section{Introduction}
\label{sec:intro}

Many stochastic phenomena are modelled by an observable process whose parameters depend on a time-changing unobserved regime, commonly modelled as a finite state-space Markov process. If the model allows for serial dependence after conditioning on the regime, we speak of a \emph{Markov switching model}. Such models are most common in economics and finance, where they were first proposed to infer business cycles from GDP growth data \citep{hamilton1989new}. Many other economic time series exhibit cyclical regime shifts, such as exchange rates \citep{engel1990long}, interest rates \citep{cai1994markov}, stock prices \citep{hamilton1994autoregressive}, commodity prices \citep{fong2002markov} and energy prices \citep{mount2006predicting}. Regime switching processes also lend themselves to modelling structural breaks in economic regimes, such as in \citep{kim1999has, mcconnell2000output}. While discrete time models are dominant in econometrics, Markov switching models also have a natural continuous time formulation as \emph{Markov switching diffusions}, i.e. diffusion processes whose drift and volatility functions change according to a continuous time Markov jump process. For example, \citep{yan2014moving} apply a driftless switching model to the task of animal tracking. Moreover, mathematicians have long investigated stability and optimal control of such models \citep{ghosh1993optimal, basak1996stability, mao2006stochastic}. Conversely, since the transition law of most diffusion models is not analytically available, likelihood-based inference is not immediately possible.

\begin{figure}
  \centering
  \includegraphics[scale=.66]{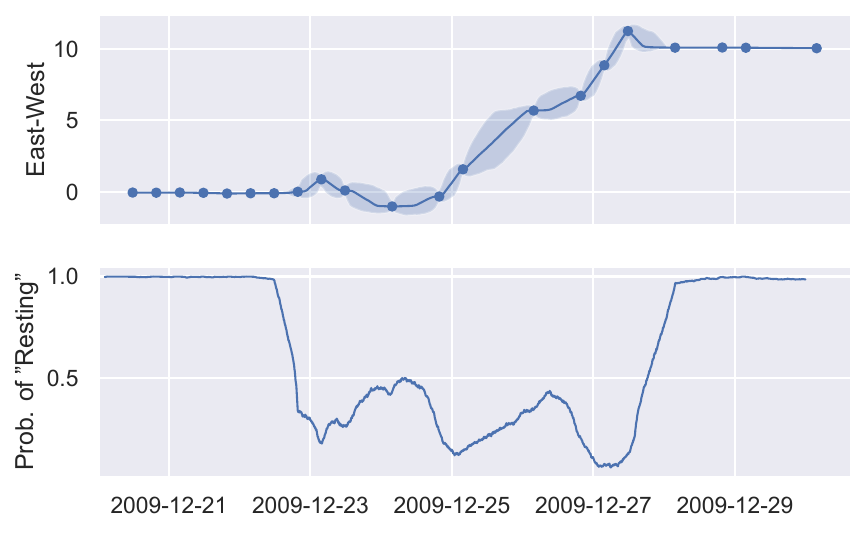}
  \caption{[Mountain Lion tracking model] Top: Markers show East-West position measurements over 10 days in 2009, the line is the predictive median, and the shaded area the 75\% prediction interval. Bottom: marginal posterior probabilities for the ``resting'' regime.}
  \label{fig:f109-mcmc-latent}
\end{figure}

As exemplified by \citep{hamilton1989new}, most of the existing literature on Markov switching uses discrete time models, thereby avoiding the technical challenges of the continuous time setting. Others seek to address the problem by either restricting the diffusion process to an analytically tractable family \citep{blackwell2003bayesian, hahn2010markov}, or by using an Euler-Maruyama-type discrete time approximation of the process dynamics \citep{hodgson1999bayesian, liechty2001markov}. Higher-order approximation schemes were proposed by \citep{ait2002maximum} and applied to Markov switching diffusions in \citep{choi2009regime}. Since such approximations are biased, an MCMC algorithm that relies on them yields samples from an approximate posterior. As the Euler-type discretization is refined, the approximation to the posterior improves consistently. Nonetheless, whereas Monte Carlo error control is well-studied and understood, the discretization bias resulting from Euler approximation is difficult to quantify in any given finite sample setting, and more so in the presence of regime switching. Asymptotically and in the instance of Ito diffusions, the error of Euler-type schemes in estimating test functions is of order $\class{O}(\text{effort}^{-1/3})$ \citep{duffie1995efficient}, which in special cases can be reduced to $\class{O}(\text{effort}^{-1/2})$ \citep{giles2008multilevel}. Accordingly, wherever feasible, we deem it preferable in terms of asymptotic efficiency and transparency to apply \emph{exact} methods, understood as being subject to Monte Carlo error only. This applies in particular to the Markov switching context, where approximation error is less understood, see e.g. \citep{mao2007approximations}, and which presents particular problems due to off-equilibrium effects upon a change of latent regime, where drift can be large and highly variable. In addition, by preserving the full continuous-time setting, our solution provides exact inference for both the latent and the observable process along their entire continuous domain - see Figure \ref{fig:f109-mcmc-latent}, which pertains to the animal tracking application explored in Section \ref{sec:demo}. In particular, we observe that predictive intervals automatically account for periods of low (``resting'') and high variability (``moving'') of observations. In the same section, we also empirically examine the extent of the bias when estimating the tracking model with an approximate algorithm. 

In that light, we will construct a Markov Chain with stationary distribution corresponding to the \emph{exact} posterior. As a result, posterior summaries are subject to Monte Carlo estimation error only, and if the \emph{Monte Carlo Central Limit Theorem} holds, we recover $\class{O}(\text{effort}^{-1/2})$ asymptotics. This leverages an extensive literature on exact simulation of Ito diffusions \citep{beskos2008factorisation} as well as Bayesian posterior simulation for Ito diffusions \citep{gonccalves2017barker} and jump diffusions \citep{gonccalves2023exact}. Those methods incorporate the missing, infinite-dimensional diffusion paths in order to exploit the \emph{complete diffusion likelihood}. Critically, even though the algorithm targets a distribution on the space of infinite-dimensional paths, it only requires the evaluation of the diffusion path on a finite subset of times, which is extended as required for the propagation of the algorithm by interpolating the previously revealed path skeleton. This is known as \emph{retrospective simulation}. While various retrospective techniques have been applied to construct exact MCMC algorithms, we focus on the \emph{Bernoulli factory MCMC} approach, seen e.g. in \citep{gonccalves2017barker, gonccalves2023exact}, which allows for the implementation of accept/reject coin flips without explicitly evaluating the potentially intractable probability of acceptance. Relative to other approaches to constructing exact MCMC algorithms, such as pseudo-marginal MCMC, this has the benefit of minimizing the extent of data augmentation, which is liable to negatively impact Markov chain mixing and inflate estimation variance, sometimes in intransparent ways. As it happens, Bernoulli MCMC algorithms also have more transparent failure modes, in that their iteration time will inflate, unlike the deterioration in mixing that occurs e.g. in pseudo-marginal algorithms \citep{sherlock2015efficiency, andrieu2015convergence}. Moreover, in our experiments, we observe close to linear scaling of computational cost in data size. This stands in contract with pseudo-marginal algorithms, where a quadratic cost per iteration has to be incurred in order to keep the variance of the likelihood estimator constant as data size increases, as recommended e.g. by \citep{sherlock2015efficiency, doucet2015efficient}.

Though we adopt much of the framework seen in \citep{gonccalves2023exact} and leverage some innovations from \citep{vats2020efficient, stumpf2025scalable}, we surmount a range of challenges particular to the Markov switching setting, and in doing so improve on established exact methods. Those challenges arise due to the presence of the latent Markov jump process, which can result in more erratic drift and volatility patterns. It is only through careful exploitation of the model structure, flexible initialization and adaptation procedures, as well as improvements to the algorithmic complexity of Bernoulli MCMC methods, that practical MCMC algorithms can arise in this context, even for modest sample sizes. Indeed, our results demonstrate good scalability properties without any ad-hoc tuning. In doing so, we set a template for how to apply exact methods in complex latent process models.

In summary, we list our contributions as
\begin{itemize}
\item providing the first, fully fledged solution to exact Bayesian inference in Markov switching diffusion models;
\item implementing exact inference in diffusion-based models in a way that is robust to model and data, doesn't require hand-tuning or model-specific implementation, and which scales with data size in various regimes;
\item deriving a novel discretized algorithm from the exact algorithm, and examining its properties empirically.
\end{itemize}

The paper is organized as follows. Section \ref{sec:intro} continues with a presentation of the model of interest and the notational conventions. Section \ref{sec:aug} introduces the augmentation scheme that underpins our inference algorithms. Section \ref{sec:gibbs} describes an MCMC algorithm for posterior sampling, and elaborates on various underlying techniques. Section \ref{sec:map} describes an analogous MCEM algorithm for MAP estimation. Section \ref{sec:demo} demonstrates both posterior sampling and point estimation as well as approximation bias on an authentic animal tracking time series. We close with a study of the algorithm's scalability in Section \ref{sec:simstud}.

\subsection{The Markov Switching Diffusion Setting}
\label{ssec:msdiff}

The regime-switching framework in this paper is as follows. Let $Y$ be a latent, discrete space, continuous time Markov jump process with regimes $\class{Y} = \braces{1, \dots, k}$, taking values in the set $\class{K}$ of $([0, \omega] \mapsto \class{Y})$ Càdlàg functions. The jump process evolves according to its generator matrix $\lambda$, where $\lambda_{i, j \neq i} \geq 0$ are the jump rates from regime $i$ to $j$ and the diagonal elements are given by $\lambda_{ii} = -\sum_{i \neq j} \lambda_{ij}$. We follow the convention of denoting the exit rates $\lambda_{i} = -\lambda_{ii}$. The density function of $Y$ is defined with respect to the measure $\meas{L}$ induced by a rate 1 marked Poisson process. Define the Markov switching \emph{stochastic differential equation} (SDE)
\begin{equation}
  \dd{V_{t}} = \mu_{\theta}(V_{t}, Y_{t}) \dd{t} + \sigma_{\theta}(V_{t}) \rho_{\theta}(Y_{t}) \dd{W_{t}}, \qquad (V_{0} = v_{0}, \quad V_{0} \perp Y_{0})
\end{equation}
where $W$ is a standard Brownian motion and $\theta$ is a parameter vector. The methodology in this paper naturally extends to multivariate diffusions, with the caveat of stricter requirements on the functional form of $\sigma_{\theta}(V_{t})$. Suppose that the SDE admits a unique solution for every $y \in \class{K}$ and $\theta$ and therefore a Markov transition density $\pi(v_{t + \epsilon} | v_{t}, y, \theta)$. Assume that $V$ with state space $\class{V}$ is observed at times $0, s_{1}, s_{2}, \dots$ with values $v_{0}, v_{s_{1}}, v_{s_{2}}, \dots$ while all other quantities are unknown, i.e. $\theta$ and $\lambda$ denote realizations of the random variables $\Theta$ and $\Lambda$.  The factorization of the volatility term need not be unique and this arbitrariness does not affect the algorithm presented in the paper.  The overarching goal is to devise a Markov chain Monte Carlo algorithm that targets the posterior
\begin{equation}
  \pi(y, \theta, \lambda | v_{0}, v_{s_{1}}, v_{s_{2}}, \dots) \propto \pi(y | \lambda) \pi(\theta) \pi(\lambda) \pi(v_{s_{1}} | v_{0}, y, \theta) \pi(v_{s_{2}} | v_{s_{1}}, y, \theta) \cdots
\end{equation}
for a given product prior $\pi(\theta, \lambda) = \pi(\theta) \pi(\lambda)$. Except for rare special cases, the conditional transition density $\pi(v_{t + \epsilon} | v_{t}, y, \theta)$ is intractable. Hence, standard methods of likelihood-based inference are not directly applicable. Therein lies the fundamental challenge to inference in continuous time models.

\subsection{Conventions}
\label{ssec:conv}

Throughout the paper, random variables will be written in uppercase letters and realizations thereof in lower case. Moreover, for a random variable $A$ and realization $a$, $\pi(a)$ refers to the density at $a$ if $A$ is continuous and the probability mass at $a$ if $A$ is discrete. For a probability measure $\meas{M}$ on $(A, B)$, $\meas{M}_{|a}$ denotes the conditional measure on $B$ after observing $\braces{A = a}$. For three sets $a, b, c$ such that $a \subseteq b$ and a function $f: b \to c$, $f(a)$ denotes $\braces{f(\dot{a}): \dot{a} \in a}$. Analogously, given a continuous path $v$ and a set of times $s$, we use the notation $v_{s}$ to refer to $\braces{v_{s}: \dot{s} \in s}$. For an ordered set $s = \braces{s_{0} < s_{1} < \dots}$, $\braces{(\dot{s} \sim \ddot{s}) \in s}$ refers to the set of paired neighbors $(s_{0}, s_{1}), (s_{1}, s_{2}), \dots$, and we use $\omega$ to denote the ``end of time''.

\section{Data Augmentation Strategy}
\label{sec:aug}

In this section, we derive the complete transition density of the model with respect to an appropriate dominating measure, amenable to efficient Gibbs sampling. Since we presume that the transition density is intractable, it is not possible to target the marginal posterior over $(Y, \Theta, \Lambda)$ with standard methods. We follow the common strategy of augmenting the state space with the missing diffusion bridges $V_{(\dot{s}, \ddot{s})}$ in between the observation pairs $(\dot{s} \sim \ddot{s}) \in s$. This results in a posterior distribution with convenient conditional independence structure. In fact, it is useful to segment the diffusion path according to the union of observations $V_{s}$ and imputed values $V_{R} = v_{r}$, where $R$ is the set of times where $Y$ changes values. Defining $\tau = s \cup r$, this yields the complete likelihood
\begin{equation}
  \pi(v_{(0, \omega]} | v_{0}, y, \theta) \propto \prod_{(\dot{s} \sim \ddot{s}) \in s} \pi(v_{(\dot{s}, \ddot{s}]} | v_{\dot{s}}, y_{(\dot{s}, \ddot{s}]}, \theta) = \prod_{(\dot{\tau} \sim \ddot{\tau}) \in \tau} \pi(v_{(\dot{\tau}, \ddot{\tau}]} | v_{\dot{\tau}}, y_{\dot{\tau}}, \theta)
\end{equation}
with respect to an appropriate dominating measure. Such a \emph{centered parameterization} is not amenable to Gibbs sampling because the full conditionals $\pi(v_{[0, \omega] \setdiff s} | v_{s}, y, \theta)$ and $\pi(\theta | v, y)$ are mutually singular for distinct values $\theta$ and $\theta^{\dagger}$ \citep{roberts2001inference}. For instance, the quadratic variation increment $\dd{\angls{V}_{t}} = \sigma_{\theta}^{2}(V_{t}) \rho_{\theta}^{2}(Y_{t}) \dd{t}$ is a deterministic function of $V$, $Y$ and $\theta$, so $\pi(\theta | v, y)$ can only assign positive probability to values of $\theta$ for which $\angls{V}_{t}$ is preserved. The standard solution consists of changing variables to a \emph{non-centered parameterization}, where the diffusion bridges and $\theta$ are a priori independent. Equivalently, the change of variables results in a density with respect to a fixed dominating measure. To carry out this change of variables, we define the \emph{Lamperti transform}
\begin{equation}
  \eta_{\theta}(a) = \int_{v^{*}}^{a} \frac{\dd{b}}{\sigma_{\theta}(b)}, \qquad (v^{*}, a \in \class{V})
\end{equation}
which transforms $V$ to a process with volatility coefficient $\rho_{\theta}(Y_{t})$. Consequently, under regularity conditions and as a consequence of Girsanov's Theorem, the transformed process is absolutely continuous with respect to a simple dominating process, e.g. scaled Brownian motion. The Lamperti transform exists under mild conditions on $\sigma_{\theta}$ in the univariate diffusion setting, though conditions are more stringent in the multivariate setting. For $X_{t} = \eta_{\theta}(V_{t})$, we notice that the endpoints of $X_{(\dot{\tau}, \ddot{\tau})}$ are still determined by $\theta$ through $\eta_{\theta}$. We complete the re-parameterization by defining
\begin{equation}
  \zeta_{\theta}(x_{t}; y_{\dot{\tau}}, v_{\braces{\dot{\tau}, \ddot{\tau}}}) = \braces*{x_{t} - \eta_{\theta}(v_{\dot{\tau}}) - (\eta_{\theta}(v_{\ddot{\tau}}) - \eta_{\theta}(v_{\dot{\tau}})) \frac{t - \dot{\tau}}{\ddot{\tau} - \dot{\tau}}} / \rho_{\theta}(y_{\dot{\tau}}), \qquad (t \in (\dot{\tau}, \ddot{\tau}))
\end{equation}
which transforms $Z_{(\dot{\tau}, \ddot{\tau})} = \zeta_{\theta}(X_{(\dot{\tau}, \ddot{\tau})}; y_{\dot{\tau}}, v_{\braces{\dot{\tau}, \ddot{\tau}}})$ into a diffusion bridge with endpoints at 0. We also define $\zeta_{\theta}^{-1}$ as the inverse of $\zeta_{\theta}$ in the first argument:
\begin{equation}
  \zeta_{\theta}^{-1}(z_{t}; y_{\dot{\tau}}, v_{\braces{\dot{\tau}, \ddot{\tau}}}) = \eta_{\theta}(v_{\dot{\tau}}) + (\eta_{\theta}(v_{\ddot{\tau}}) - \eta_{\theta}(v_{\dot{\tau}})) \frac{t - \dot{\tau}}{\ddot{\tau} - \dot{\tau}} + \rho_{\theta}(y_{\dot{\tau}}) z_{t}. \qquad (t \in (\dot{\tau}, \ddot{\tau}))
\end{equation}
The following proposition establishes that parameterizing the state space in terms of $Z_{(\dot{\tau}, \ddot{\tau})}$ is appropriate.

\begin{proposition}[Non-centered augmentation for Markov switching diffusions]
  \label{prop:aug}
  Let $V$ have a unique solution for every $y$ and $\theta$, and assume that
  \begin{itemize}
  \item $\eta_{\theta}$ exists, and $\delta_{\theta}(a, b) = \braces{\mu_{\theta}(\cdot, b) / \sigma_{\theta} - \sigma_{\theta}'/2} \circ \eta_{\theta}^{-1}(a)$ is continuously differentiable in $a$ on $\class{V}$.
  \item The \emph{Novikov condition} applies, i.e. $\op{E}[X_{(0, \omega]}]{\exp\braces*{\int_{0}^{\omega} \delta_{\theta}^{2}(X_{t}, b) \dd{t}} | \braces{Y_{(0, \omega]} = b}, x_{0}, \theta} < \infty$ for every $\theta, b \in \class{Y}, x_{0} \in \eta_{\theta}(\class{V}), \omega < \infty$.
  \end{itemize}
  On that basis, define
  \begin{gather}
    \varphi_{\theta}(a, b) = \frac{1}{2} \parens*{\frac{\delta_{\theta}^{2}(a, b)}{\rho_{\theta}^{2}(b)} + \partial_{a} \delta_{\theta}(a, b)}, \\
    \Delta_{\theta}(a, b) = \int \delta_{\theta}(a, b) \dd{a}, \\
    \begin{aligned}
    \label{eq:tractable}
    h_{\theta}(y_{\dot{\tau}}, v_{\braces{\dot{\tau}, \ddot{\tau}}}) 
    & = \bars{\eta_{\theta}'(v_{\ddot{\tau}})} \op{N}{\eta_{\theta}(v_{\ddot{\tau}}); \eta_{\theta}(v_{\dot{\tau}}), (\ddot{\tau} - \dot{\tau})  \rho_{\theta}^{2}(y_{\dot{\tau}})} \\
    & \quad \times \exp\braces*{\rho_{\theta}^{-2}(y_{\dot{\tau}}) \braces{ \Delta_{\theta}(\eta_{\theta}(v_{\ddot{\tau}}), y_{\dot{\tau}}) - \Delta_{\theta}(\eta_{\theta}(v_{\dot{\tau}}), y_{\dot{\tau}})}}.
    \end{aligned}
  \end{gather}
  Then,
  \begin{equation}
    \label{eq:augtrans}
    \pi(z_{(\dot{\tau},\ddot{\tau})}, v_{\ddot{\tau}} | v_{\dot{\tau}}, y_{\dot{\tau}}, \theta)
    = h_{\theta}(y_{\dot{\tau}}, v_{\braces{\dot{\tau}, \ddot{\tau}}}) \exp\braces*{-\int_{\dot{\tau}}^{\ddot{\tau}} \varphi_{\theta}(\zeta_{\theta}^{-1}(z_{t}; y_{\dot{\tau}}, v_{\braces{\dot{\tau}, \ddot{\tau}}}), y_{\dot{\tau}}) \dd{t}}
  \end{equation}
  is a density with respect to $\meas{B}_{(\dot{\tau}, \ddot{\tau})} \times \leb$, where $\meas{B}_{(\dot{\tau}, \ddot{\tau})}$ is the Brownian bridge measure conditioned on hitting 0 at times $\dot{\tau}$ and $\ddot{\tau}$, and it satisfies
  \begin{equation}
    \int \pi(z_{(\dot{\tau},\ddot{\tau})}, v_{\ddot{\tau}} | v_{\dot{\tau}}, y_{\dot{\tau}}, \theta) \meas{B}_{(\dot{\tau}, \ddot{\tau})}(\dd{z_{(\dot{\tau},\ddot{\tau})}}) = \pi(v_{\ddot{\tau}} | v_{\dot{\tau}}, y_{\dot{\tau}}, \theta).
  \end{equation}
  \begin{proof}
    See Supplement \ref{sec:proofaug}.
  \end{proof}
\end{proposition}

Accordingly, the non-centered complete likelihood
\begin{equation}
  \begin{split}
    \pi(v_{\tau \setdiff \braces{0}}, z | v_{0}, y, \theta) = \prod_{(\dot{\tau} \sim \ddot{\tau}) \in \tau} \pi(z_{(\dot{\tau},\ddot{\tau})}, v_{\ddot{\tau}} | v_{\dot{\tau}}, y_{\dot{\tau}}, \theta)
  \end{split}
\end{equation}
marginalizes to $\pi(v_{s \setdiff \braces{0}} | v_{0}, y, \theta)$ by Proposition \ref{prop:aug}, and its dominating measure is suitably invariant in $\theta$. Note that $\pi(v_{\tau \setdiff \braces{0}}, z | v_{0}, y, \theta)$ cannot be directly evaluated due to the presence of a time integral over $z$. Indeed, $z$ cannot even be exhaustively stored in memory. Nonetheless, the algorithm presented in the next section is for all intents and purposes an MCMC algorithm on an infinite-dimensional state space, so we mostly treat the actual representation of $z$ as an implementation detail, and our notation refers to the full path $z$, unless that implementation is relevant.

\section{Exact Posterior Sampling using Barker-within-Gibbs}
\label{sec:gibbs}

We now present a Gibbs sampler that targets the augmented posterior $\pi(v_{r}, z, y, \theta, \lambda | v_{s})$ by way of the full conditionals
\begin{align}
  (\Theta, \Lambda): \quad & \pi(\theta, \lambda | v_{\tau}, z, y) \propto \pi(\theta) \pi(\lambda) \prod_{(\dot{\tau} \sim \ddot{\tau}) \in \tau} \pi(z_{(\dot{\tau}, \ddot{\tau})}, v_{\ddot{\tau}} | v_{\dot{\tau}}, y_{\dot{\tau}}, \theta) \pi(y_{\ddot{\tau}} | y_{\dot{\tau}}, \lambda), \\
  (V_{R}, Z, Y): \quad & \pi(v_{r}, z, y | v_{s}, \theta, \lambda) \propto \prod_{(\dot{\tau} \sim \ddot{\tau}) \in \tau} \pi(z_{(\dot{\tau}, \ddot{\tau})}, v_{\ddot{\tau}} | v_{\dot{\tau}}, y_{\dot{\tau}}, \theta) \pi(y_{\ddot{\tau}} | y_{\dot{\tau}}, \lambda).
\end{align}
Keeping in mind that the set of jumps $R$ and event times $T$ follow deterministically from $Y$, and that $R$ is almost surely finite, the dominating measure of the second full conditional is the product measure $\meas{L}(\dd{y}) \prod_{(\dot{\tau} \sim \ddot{\tau}) \in \tau} \meas{B}_{(\dot{\tau}, \ddot{\tau})}(\dd{z_{(\dot{\tau}, \ddot{\tau})}}) \prod_{\dot{r} \in r} \leb(\dd{v_{\dot{r}}})$, which is invariant in $\theta$. This blocking offers various opportunities for exploiting conditional independence and tuning proposals. In fact, the $(\Theta, \Lambda)$-update decomposes into the independent updates
\begin{align}
  \Theta: \quad & \pi(\theta | v_{\tau}, z, y) \propto \pi(\theta) \prod_{(\dot{\tau} \sim \ddot{\tau}) \in \tau} \pi(z_{(\dot{\tau}, \ddot{\tau})}, v_{\ddot{\tau}} | v_{\dot{\tau}}, y_{\dot{\tau}}, \theta), \\
  \Lambda: \quad & \pi(\lambda | y) \propto \pi(\lambda) \prod_{(\dot{r} \sim \ddot{r}) \in r} \pi(y_{\ddot{r}} | y_{\dot{r}}, \lambda).
\end{align}
Further simplifications arise if the SDE depends on a separate set of parameters for each latent mode. In that case, the updates to those parameters may also be carried out independently. We exploit that structure in the models seen in Sections \ref{sec:demo} and \ref{sec:simstud}. 

Each of the Gibbs updates poses distinct challenges. For the $(V_{R}, Z, Y)$-update, the construction and simulation of the proposal is fairly intricate, and the accept-reject procedure is a Bernoulli factory that implements \emph{Barker's algorithm} within Gibbs \citep{barker1965monte}, which just as Metropolis-within-Gibbs leaves the full conditional distribution invariant. We introduce the Bernoulli factory approach to diffusion inference in Section \ref{ssec:retro} before specifying the proposal and the factory for the $(V_{R}, Z, Y)$-update in Section \ref{ssec:retro}. The $\Theta$-update again relies on a Barker-within-Gibbs update with a simple random walk proposal, though specific scalability challenges arise when using Bernoulli factories in such an instance. We describe and address those challenges in Section \ref{ssec:param}. Finally, the $\Lambda$-update is conjugate for the class of priors discussed in Section \ref{ssec:gen}, and may thus be sampled exactly from the full conditional. We close the Section with a discussion of various practical implementation challenges in Section \ref{ssec:pracem}.

\subsection{Retrospective Simulation for Diffusion Inference}
\label{ssec:retro}

In what follows, both updates of the Gibbs sampler are carried out by accept-reject coin flips, where the acceptance probability depends on the intractable complete likelihood $\pi(v_{\tau \setdiff \braces{0}}, z | v_{0}, y, \theta)$. As originally suggested in the Ito diffusion context by \citep{gonccalves2017barker} and later applied to jump diffusions in \citep{gonccalves2023exact}, such evaluations can be entirely avoided without affecting the dynamics of the Markov chain. This is accomplished through the conjunction of three insights: Firstly, we may construct events with probabilities proportional to $\pi(v_{\tau \setdiff \braces{0}}, z | v_{0}, y, \theta)$ that depend only on a finite number of evaluations of $z$. In the instance of diffusions, this is known as the \emph{Poisson coin algorithm} \citep{beskos2006retrospective}. Secondly, $z$ may be left indeterminate until specific evaluations are needed to propagate the Markov chain. This \emph{skeleton} of $z$ is then simulated \emph{retrospectively} at the required locations. Finally, using \emph{Bernoulli factory} techniques, we may construct an accept-reject coin flip from Poisson coin flips.

\begin{figure}
  \centering
  \begin{tikzpicture}[edge from parent/.style={draw,latex-}, level 1/.style={level distance=15mm}, level 2/.style={level distance=15mm, sibling distance=80mm}]
  \node (start) at (0, 0) {\textbf{start}}; 
  \node (f0) at (0, -1) {$C_{0}$};
  \draw[->] (start) -- (f0);
  \node (f1) at (-2, -2) {$C_{1}$};
  \draw[->, bend right] (f0) to node[midway, above left] {$\frac{c_{1}}{c_{1} + c_{2}}$} (f1);
  \draw[->, bend right] (f1) to node[very near start, below right] {$1 - p_{1}$} (f0);
  \node (f2) at (2, -2) {$C_{2}$};
  \draw[->, bend left] (f0) to node[midway, above right] {$\frac{c_{2}}{c_{1} + c_{2}}$} (f2);
  \draw[->, bend left] (f2) to node[very near start, below left] {$1 - p_{2}$} (f0);
  \node (r1) at (-4.5, -2) {\textbf{return} 1};
  \draw[->] (f1) -- node[midway, below] {$p_{1}$} (r1);
  \node (r0) at (4.5, -2) {\textbf{return} 0};
  \draw[->] (f2) -- node[midway, below] {$p_{2}$} (r0);
  \end{tikzpicture}
  \caption{Probability flow diagram of the vanilla 2-coin algorithm. Nodes ($C_{0}$, $C_{1}$, $C_{2}$) refer to coin flips, edges give the probabilities of moving to the corresponding node.}
  \label{fig:2c}
\end{figure}
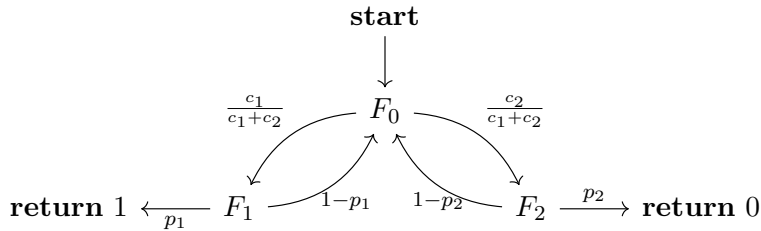

Before delving deeper into the Bernoulli MCMC approach, we briefly recall some fundamentals of accept-reject MCMC. Suppose we are targeting the density $\pi(a)$ with $A$ taking values in the state space $\class{A}$ and a \emph{proposal density} $\kappa(a^{\dagger} | a)$. The Metropolis-Hastings (MH) acceptance probability for a proposal $a^{\dagger}$ is $\alpha_{\text{MH}}(a, a^{\dagger}) = 1 \land \braces{\pi(a^{\dagger}) \kappa(a | a^{\dagger}) / \braces{\pi(a) \kappa(a^{\dagger} | a)}}$, and its repeated application to random proposals results in a Markov chain with stationary density $\pi(a)$ under mild conditions on $\kappa$ due to the \emph{detailed balance} or \emph{reversibility} property. It may also be used to update a full conditional distribution within a Gibbs sampler when direct sampling thereof is not possible, which is known as \emph{Metropolis-within-Gibbs}. There are other acceptance probabilities that preserve detailed balance, such as accepting with odds $\alpha_{\text{B}}(a, a^{\dagger}) / \alpha_{\text{B}}(a^{\dagger}, a) = \pi(a^{\dagger}) \kappa(a | a^{\dagger}) / \braces{\pi(a) \kappa(a^{\dagger} | a)}$, as proposed by \citep{barker1965monte}. These are rarely applied since MH induces smaller autocorrelation in the Markov chain than other reversible acceptance probabilities, as shown in \citep{peskun1973optimum}. Nonetheless, Barker's acceptance probability takes on a special role in the intractable likelihood context. Firstly, as shown in \citep{latuszynski2013clts}, when estimating a test function $f(a)$ by its ergodic average under each Markov chain, the asymptotic variances of the two estimators are within a factor of 2, so any problem that may be addressed by MH is also solvable with Barker. Secondly, the Barker accept-reject coin is more amenable to being constructed through a \emph{Bernoulli factory}. To expand on the latter, suppose that $\alpha_{\text{B}}$ admits the factorization
\begin{equation}
  \label{eq:2cfact}
  \frac{\alpha_{\text{B}}(a, a^{\dagger})}{\alpha_{\text{B}}(a^{\dagger}, a)} = \frac{\pi(a^{\dagger}) \kappa(a^{\dagger} | a)}{\pi(a) \kappa(a | a^{\dagger})} = \frac{c_{1} p_{1}}{c_{2} p_{2}},
\end{equation}
where $c_{1}, c_{2}$ are tractable constants and $p_{1}, p_{2} \in [0, 1]$ are potentially intractable, but for which we can simulate coins with equivalent probability, e.g. by way of an unbiased estimator. Then, provided a stream of coin flips of heads-probability  $p_{1}$ and another of heads-probability $p_{2}$, the \emph{2-coin algorithm} generates coins of probability $\alpha_{\text{B}}(a, a^{\dagger})$ \citep{gonccalves2017barker}. We summarize its operation in Figure \ref{fig:2c}. Then, on a high level, the Bernoulli MCMC algorithm generates $p_{1}$- and $p_{2}$-coins through the Poisson coin method, which we summarize in Supplement \ref{sec:poisson}, revealing the diffusion path retrospectively to the resolution required for the coin flips. On that basis, the $\alpha_{\text{B}}$-coins are simulated by the 2-coin algorithm, without ever evaluating $\alpha_{\text{B}}$. In fact,  the evaluation of $\alpha_{\text{B}}$ is entirely ancillary, since it is merely used to determine the binary event $\braces{\alpha_{\text{B}}(a, a^{\dagger}) > U}$ for $U \sim \op{Uniform}{0, 1}$. How the accept-reject coin is constructed does not affect the dynamics of the Markov chain. On the other hand, the runtime of the 2-coin algorithm is highly dependent on the chosen factorization: the number of loops in the 2-coin algorithm is a geometric random variable with expectation
\begin{equation}
  \label{eq:2ctime}
  \frac{c_{1} + c_{2}}{c_{1} p_{1} + c_{2} p_{2}},
\end{equation}
which diverges as $p_{1}, p_{2} \to 0$. This is critical to the $\theta$-update described in Section \ref{ssec:param}, and extensions to the standard 2-coin approach are required to prevent the computational cost from diverging exponentially as $\omega$ increases. Of particular relevance is the 
\emph{divide-and-conquer 2-coin algorithm}, proposed by \citep{stumpf2025scalable} and applied in Section \ref{ssec:param}, and the Portkey 2-coin algorithm, proposed by \citep{vats2020efficient} and applied to the experiments in Sections \ref{sec:demo} and \ref{sec:simstud}.

In order to simulate $p_{1}$- and $p_{2}$-coins in the broadest set of Markov switching models, the skeleton $\breve{z}_{(\dot{\tau}, \ddot{\tau})}$ of the bridge $z_{(\dot{\tau}, \ddot{\tau})}$ must satisfy three requirements. Firstly, it must incorporate all previously evaluated locations on the path, such that all retrospective simulations are mutually consistent. Secondly, it must provide lower and upper bounds
\begin{equation}
  -\infty < \lb{z}_{(\dot{\tau}, \ddot{\tau})} \leq z_{t} \leq \ub{z}_{(\dot{\tau}, \ddot{\tau})} < \infty, \qquad (t \in (\dot{\tau}, \ddot{\tau}))
\end{equation}
which can then be propagated to bounds on $\tilde{\varphi}_{\theta}(z_{t}, y_{\dot{\tau}}, v_{\braces{\dot{\tau}, \ddot{\tau}}}) = \varphi_{\theta}(\zeta_{\theta}^{-1}(z_{t}; y_{\dot{\tau}}, v_{\braces{\dot{\tau}, \ddot{\tau}}}), y_{\dot{\tau}})$. These are necessary to bound the time integral $\int_{\dot{\tau}}^{\ddot{\tau}} \tilde{\varphi}_{\theta}(z_{t}, y_{\dot{\tau}}, v_{\braces{\dot{\tau}, \ddot{\tau}}}) \dd{t}$ and thereby the posterior odds, which is a prerequisite to constructing the factorization in \eqref{eq:2cfact}. An upper bound on the time integral is also necessary for unbiased estimation of the intractable coin probabilities. Thirdly, it must be possible to recursively simulate $z_{(\dot{\tau}, \ddot{\tau})}$ at an additional finite set of times according to the initial proposal distribution, conditional on the skeleton $\breve{z}_{(\dot{\tau}, \ddot{\tau})}$. In this paper, the proposal distribution will always be the dominating measure $\meas{B}_{(\dot{\tau}, \ddot{\tau})}$. Modulo these requirements, any method may in principle be used to generate the coins. The most general-purpose representations we're aware of are  the Bessel layer representation by \cite{beskos2008factorisation} and the $\epsilon$-strong layer representation by \cite{pollock2016exact}. Both representations rely on restricting the state space of $z$ to a random compact subset of $\reals$, with the latter being a richer representation, but the former being more numerically reliable and easier to implement. Therefore, we use the Bessel layer representation in practice, as seen for example in \cite{gonccalves2023exact}.


In what follows, we merely presume that some lower and upper bounds $\lb{z}_{(\dot{\tau}, \ddot{\tau})}$ and $\ub{z}_{(\dot{\tau}, \ddot{\tau})}$ are available. Defining
\begin{equation}
  \class{X}_{(\dot{\tau}, \ddot{\tau})} = [\rho_{\theta}(y_{\dot{\tau}}) \lb{z}_{(\dot{\tau}, \ddot{\tau})} + \eta_{\theta}(v_{\dot{\tau}}) \land \eta_{\theta}(v_{\ddot{\tau}}),  \rho_{\theta}(y_{\dot{\tau}}) \ub{z}_{(\dot{\tau}, \ddot{\tau})} + \eta_{\theta}(v_{\dot{\tau}}) \lor \eta_{\theta}(v_{\ddot{\tau}})]
\end{equation}
and observing that $x_{t} \in \class{X}_{(\dot{\tau}, \ddot{\tau})}$, we may define
\begin{gather}
  \lb{\tilde{\varphi}}_{\theta}(z_{(\dot{\tau}, \ddot{\tau})}, y_{\dot{\tau}}, v_{\braces{\dot{\tau}, \ddot{\tau}}}) = \inf_{a \in \class{X}_{(\dot{\tau}, \ddot{\tau})}} \varphi_{\theta}(a, y_{\dot{\tau}}), \quad \ub{\tilde{\varphi}}_{\theta}(z_{(\dot{\tau}, \ddot{\tau})}, y_{\dot{\tau}}, v_{\braces{\dot{\tau}, \ddot{\tau}}}) = \sup_{a \in \class{X}_{(\dot{\tau}, \ddot{\tau})}} \varphi_{\theta}(a, y_{\dot{\tau}}),
\end{gather}
such that $\lb{\tilde{\varphi}}_{\theta}(z_{(\dot{\tau}, \ddot{\tau})}, y_{\dot{\tau}}, v_{\braces{\dot{\tau}, \ddot{\tau}}}) \leq \tilde{\varphi}_{\theta}(z_{t}, y_{\dot{\tau}}, v_{\braces{\dot{\tau}, \ddot{\tau}}}) \leq \ub{\tilde{\varphi}}_{\theta}(z_{(\dot{\tau}, \ddot{\tau})}, y_{\dot{\tau}}, v_{\braces{\dot{\tau}, \ddot{\tau}}})$. Solving for the infimum and supremum (or bounds thereon) is a model-specific task, but may be done automatically by symbolic algebra packages.

\subsection{Hidden Data Update}
\label{ssec:hidden}

To begin with, we consider an independence proposal for updating $\pi(v_{r}, z, y | v_{s}, \theta, \lambda)$. This is sufficient when $v_{s}$ is fairly uninformative, or the time horizon short, and results in notation that is easier to parse. Failing that, proposals can be localized, as described in Supplement \ref{sec:localize}, which we apply in our computer experiments in Sections \ref{sec:demo} and \ref{sec:simstud}. We construct a Barker-within Gibbs update with the hierarchical proposal
\begin{equation}
  \kappa(v_{r^{\dagger}}^{\dagger}, z^{\dagger}, y^{\dagger} | v_{s}) \propto \kappa(y^{\dagger}) \kappa(v_{r}^{\dagger} | v_{s}, y^{\dagger}) \prod_{(\dot{\tau} \sim \ddot{\tau}) \in \tau} \kappa(z_{(\dot{\tau}, \ddot{\tau})}^{\dagger}),
\end{equation}
where $Z_{(\dot{\tau}, \ddot{\tau})}^{\dagger} \sim \meas{B}_{(\dot{\tau}, \ddot{\tau})}$ and $\kappa(z_{(\dot{\tau}, \ddot{\tau})}^{\dagger}) = 1$, with respect to $\meas{B}_{(\dot{\tau}, \ddot{\tau})}$. Using a Brownian bridge proposal ensures that the path skeletons can be simulated and extended, e.g. by the EA algorithm. $Y^{\dagger}$ is proposed independently from its prior distribution, i.e. $\kappa(y^{\dagger}) = \pi(y^{\dagger} | \lambda)$. This is simply the forward measure of the jump process, and easily simulated from as in \citep{hobolth2009simulation}. Given $Y^{\dagger}$, the proposal for $V_{r^{\dagger}}^{\dagger}$ is most readily understood in terms of $X_{r^{\dagger}}^{\dagger} = \eta_{\theta}(V_{r^{\dagger}}^{\dagger})$. We propose $X_{r^{\dagger}}^{\dagger}$ according to the dominating SDE $\dd{X}_{t} = \rho_{\theta}(Y_{t}^{\dagger}) \dd{W}_{t}$ with induced conditional measure $\meas{M}|(x_{s}, y^{\dagger}, \theta)$. By the Markov property,
\begin{equation}
\begin{split}
  \meas{M}|(x_{s}, y, \theta)(\dd{x_{r}})
  & = \prod_{(\dot{s} \sim \ddot{s}) \in s} \meas{M}|(x_{\braces{\dot{s}, \ddot{s}}}, y_{[\dot{s}, \ddot{s}]}, \theta)(\dd{x_{r \cap (\dot{s}, \ddot{s})}}) \\
  & = \prod_{(\dot{s} \sim \ddot{s}) \in s} \frac{\prod_{(\dot{\tau} \sim \ddot{\tau}) \in \tau \cap [\dot{s}, \ddot{s}]} \meas{M}|(x_{\dot{\tau}}, y_{\dot{\tau}}, \theta)(\dd{x_{\ddot{\tau}}})}{\meas{M}|(x_{\dot{s}}, y_{[\dot{s}, \ddot{s})}, \theta)(\dd{x_{\ddot{s}}})},
\end{split}
\end{equation}
and each subset $X_{r \cap (\dot{s}, \ddot{s})}$ may be simulated independently. We do so by observing that by the time change representation of the stochastic integral, the transformation
\begin{equation}
  (t, x_{t}) \mapsto \parens*{\int_{\dot{s}}^{t} \rho_{\theta}^{2}(y_{u}) \dd{u}, x_{t}}
\end{equation}
maps $X$ to a unit volatility Brownian bridge connecting $(0, x_{\dot{s}}) \to (\int_{\dot{s}}^{\ddot{s}} \rho_{\theta}^{2}(y_{u}) \dd{u}, x_{\ddot{s}})$. Conversely, a sample from that bridge at time $\int_{\dot{s}}^{t} \rho_{\theta}^{2}(y_{u}) \dd{u}$ follows the proposal law of $X_{t}$. We then obtain $V_{r} = \eta_{\theta}^{-1}(X_{r})$. The measure $\meas{M}|(x_{\braces{\dot{s}, \ddot{s}}}, y_{[\dot{s}, \ddot{s}]}, \theta)(\dd{x_{r \cap (\dot{s}, \ddot{s})}})$ is Gaussian and has density
\begin{equation}
  \kappa(x_{r \cap (\dot{s}, \ddot{s})} | x_{\braces{\dot{s}, \ddot{s}}}, y_{[\dot{s}, \ddot{s}]}) = \frac{\prod_{(\dot{\tau} \sim \ddot{\tau}) \in \tau \cap [\dot{s}, \ddot{s}]} \op{N}{x_{\ddot{\tau}}^{\dagger}; x_{\dot{\tau}}, (\ddot{\tau} - \dot{\tau}) \rho_{\theta}^{2}(y_{\dot{\tau}})}}{\op{N}{x_{\ddot{s}}; x_{\dot{s}}, \sum_{(\dot{\tau} \sim \ddot{\tau}) \in \tau \cap [\dot{s}, \ddot{s}]} (\ddot{\tau} - \dot{\tau}) \rho_{\theta}^{2}(y_{\dot{\tau}})}},
\end{equation}
from which we recover the proposal density on $V_{r \cap (\dot{s}, \ddot{s})}$ by the change of variable formula:
\begin{equation}
  \label{eq:hidpropdens}
  \kappa(v_{r \cap (\dot{s}, \ddot{s})} | v_{\braces{\dot{s}, \ddot{s}}}, y_{[\dot{s}, \ddot{s}]}) = \frac{\prod_{(\dot{\tau} \sim \ddot{\tau}) \in \tau \cap [\dot{s}, \ddot{s}]} |\eta_{\theta}^{-1}(v_{\ddot{\tau}})| \op{N}{\eta_{\theta}(v_{\ddot{\tau}}); \eta_{\theta}(v_{\dot{\tau}}), (\ddot{\tau} - \dot{\tau}) \rho_{\theta}^{2}(y_{\dot{\tau}})}}{|\eta_{\theta}^{-1}(v_{\ddot{s}})| \op{N}{\eta_{\theta}(v_{\ddot{s}}); \eta_{\theta}(v_{\dot{s}}), \sum_{(\dot{\tau} \sim \ddot{\tau}) \in \tau \cap [\dot{s}, \ddot{s}]} (\ddot{\tau} - \dot{\tau}) \rho_{\theta}^{2}(y_{\dot{\tau}})}}.
\end{equation}
Thus, the proposal density on $V_{r^{\dagger}}^{\dagger}$ is given by $\kappa(v_{r^{\dagger}}^{\dagger} | v_{s}, y^{\dagger}) = \prod_{(\dot{s} \sim \ddot{s}) \in s} \kappa(v_{r^{\dagger} \cap (\dot{s}, \ddot{s})}^{\dagger} | v_{\braces{\dot{s}, \ddot{s}}}, y_{[\dot{s}, \ddot{s}]}^{\dagger})$. We spell out the step-by-step simulation routine below.

\begin{algorithm}[H]
  \caption{Algorithm for generating proposal from $\kappa(v_{r \cap (\dot{s}, \ddot{s})} | v_{\braces{\dot{s}, \ddot{s}}}, y_{[\dot{s}, \ddot{s}]})$. $\meas{W}$ denotes the Wiener measure. \label{alg:indprop}}
  \begin{algorithmic}
    \State $x_{\braces{\dot{s}, \ddot{s}}} \gets \eta_{\theta}(v_{\braces{\dot{s}, \ddot{s}}})$
    \State $u \gets \braces{\int_{\dot{s}}^{\dot{r}} \rho_{\theta}^{2}(y_{t}) \dd{t} : \dot{r} \in r \cap (\dot{s}, \ddot{s})}$
    \State $w_{u} \sim \meas{W}|(W_{0} = x_{\dot{s}}, W(\int_{\dot{s}}^{\ddot{s}} \rho_{\theta}^{2}(y_{t}) \dd{t}) = x_{\ddot{s}})$
    \State $x_{r \cap (\dot{s}, \ddot{s})} \gets w_{u}$
    \State $v_{r \cap (\dot{s}, \ddot{s})} \gets \eta_{\theta}^{-1}(x_{r \cap (\dot{s}, \ddot{s})})$
  \end{algorithmic}
\end{algorithm}


With the proposal fully specified, we define
\begin{gather}
  d_{\theta}(z_{(\dot{\tau}, \ddot{\tau})}, y_{\dot{\tau}}, v_{\braces{\dot{\tau}, \ddot{\tau}}})
  = h_{\theta}(y_{\dot{\tau}}, v_{\braces{\dot{\tau}, \ddot{\tau}}}) \exp\braces*{(\dot{\tau} - \ddot{\tau}) \lb{\tilde{\varphi}}_{\theta}(z_{(\dot{\tau}, \ddot{\tau})}, y_{\dot{\tau}}, v_{\braces{\dot{\tau}, \ddot{\tau}}})}, \\
  q_{\theta}(z_{(\dot{\tau}, \ddot{\tau})}, y_{\dot{\tau}}, v_{\braces{\dot{\tau}, \ddot{\tau}}})
  = \exp\braces*{\int_{\dot{\tau}}^{\ddot{\tau}} \lb{\tilde{\varphi}}_{\theta}(z_{(\dot{\tau}, \ddot{\tau})}, y_{\dot{\tau}}, v_{\braces{\dot{\tau}, \ddot{\tau}}}) - \tilde{\varphi}_{\theta}(z_{t}, y_{\dot{\tau}}, v_{\braces{\dot{\tau}, \ddot{\tau}}}) \dd{t}}
\end{gather}
such that $\pi(z_{(\dot{\tau}, \ddot{\tau})}, v_{\ddot{\tau}} | v_{\dot{\tau}}, y_{\dot{\tau}}, \theta^{\dagger}) = (d_{\theta} \times q_{\theta})(z_{(\dot{\tau}, \ddot{\tau})}, y_{\dot{\tau}}, v_{\braces{\dot{\tau}, \ddot{\tau}}})$, and where $q_{\theta}(z_{(\dot{\tau}, \ddot{\tau})}, y_{\dot{\tau}}, v_{\braces{\dot{\tau}, \ddot{\tau}}}) \in [0, 1]$. On that basis, the acceptance odds can be expressed as
\begin{equation}
  \begin{split}
    & \frac{\alpha_{(V_{R}, Z, Y)}(\braces{v_{r^{\dagger}}^{\dagger}, z^{\dagger}, y^{\dagger}}, \braces{v_{r}, z, y})}{\alpha_{(V_{R}, Z, Y)}(\braces{v_{r}, z, y}, \braces{v_{r^{\dagger}}^{\dagger}, z^{\dagger}, y^{\dagger}})} \\
    & = \frac{\kappa(v_{r}, z, y | v_{s})}{\kappa(v_{r^{\dagger}}^{\dagger}, z^{\dagger}, y^{\dagger} | v_{s})} \frac{\pi(v_{r^{\dagger}}^{\dagger}, z^{\dagger}, y^{\dagger} | v_{s}, \theta, \lambda)}{\pi(v_{r}, z, y | v_{s}, \theta, \lambda)} \\
    & = \frac{\kappa(v_{r} | v_{s}, y)}{\kappa(v_{r^{\dagger}}^{\dagger} | v_{s}, y^{\dagger})} \frac{\prod_{(\dot{\tau} \sim \ddot{\tau}) \in \tau^{\dagger}} \pi(z_{(\dot{\tau}, \ddot{\tau})}^{\dagger}, v_{\ddot{\tau}}^{\dagger} | v_{\dot{\tau}}^{\dagger}, y_{\dot{\tau}}^{\dagger}, \theta)}{\prod_{(\dot{\tau} \sim \ddot{\tau}) \in \tau} \pi(z_{(\dot{\tau}, \ddot{\tau})}, v_{\ddot{\tau}} | v_{\dot{\tau}}, y_{\dot{\tau}}, \theta)} \\
    & = \frac{\overbrace{\kappa(v_{r} | v_{s}, y) \prod_{(\dot{\tau} \sim \ddot{\tau}) \in \tau^{\dagger}} d_{\theta}(z_{(\dot{\tau}, \ddot{\tau})}^{\dagger}, y_{\dot{\tau}}^{\dagger}, v_{\braces{\dot{\tau}, \ddot{\tau}}}^{\dagger})}^{c_{1}}}{\underbrace{\kappa(v_{r^{\dagger}}^{\dagger} | v_{s}, y^{\dagger}) \prod_{(\dot{\tau} \sim \ddot{\tau}) \in \tau} d_{\theta}(z_{(\dot{\tau}, \ddot{\tau})}, y_{\dot{\tau}}, v_{\braces{\dot{\tau}, \ddot{\tau}}})}_{c_{2}}} \frac{\overbrace{\prod_{(\dot{\tau} \sim \ddot{\tau}) \in \tau^{\dagger}} q_{\theta}(z_{(\dot{\tau}, \ddot{\tau})}^{\dagger}, y_{\dot{\tau}}^{\dagger}, v_{\braces{\dot{\tau}, \ddot{\tau}}}^{\dagger})}^{p_{1}}}{\underbrace{\prod_{(\dot{\tau} \sim \ddot{\tau}) \in \tau} q_{\theta}(z_{(\dot{\tau}, \ddot{\tau})}, y_{\dot{\tau}}, v_{\braces{\dot{\tau}, \ddot{\tau}}})}_{p_{2}}},
  \end{split}
\end{equation}
which gives a valid 2-coin factorization. Notice that we can generate a coin with probability $p_{2} = \prod_{(\dot{\tau} \sim \ddot{\tau}) \in \tau} q_{\theta}(z_{(\dot{\tau}, \ddot{\tau})}, y_{\dot{\tau}}, v_{\braces{\dot{\tau}, \ddot{\tau}}})$ by flipping subordinate $q_{\theta}(z_{(\dot{\tau}, \ddot{\tau})}, y_{\dot{\tau}}, v_{\braces{\dot{\tau}, \ddot{\tau}}})$-coins with the Poisson coin algorithm, and returning 1 if all subordinate coin flips are 1. Conversely, we can return 0 as soon as one of the subordinate coin flips is 0. Moreover, we can split $\braces{v_{r^{\dagger}}^{\dagger}, z^{\dagger}, y^{\dagger}}$ into separate sections at $\braces{\dot{s} \in s: y_{\dot{s}} = y_{\dot{s}}^{\dagger}}$, and accept or reject those separately. We expand on that in the localized version given in Supplement \ref{sec:localize}.

\subsection{Diffusion Parameter Update}
\label{ssec:param}

We carry out the update to $\pi(\theta | v_{\tau}, z, y)$ by way of a Barker-within-Gibbs step with generic proposal density $\kappa(\theta^{\dagger} | \theta)$. Such a proposal may be adjusted adaptively, e.g. by way of \emph{Adapting Increasingly Rarely} \citep{chimisov2018air} or other adaptive MCMC methods. The proposed value $\theta^{\dagger}$ has acceptance odds
\begin{equation}
  \begin{split}
    \frac{\alpha_{\Theta}(\theta, \theta^{\dagger})}{\alpha_{\Theta}(\theta^{\dagger}, \theta)}
    & = \frac{\kappa(\theta | \theta^{\dagger})}{\kappa(\theta^{\dagger} | \theta)} \frac{\pi(\theta^{\dagger} | v_{\tau}, z, y)}{\pi(\theta | v_{\tau}, z, y)}
    = \frac{\kappa(\theta | \theta^{\dagger})}{\kappa(\theta^{\dagger} | \theta)} \frac{\pi(\theta^{\dagger})}{\pi(\theta)} \prod_{(\dot{\tau} \sim \ddot{\tau}) \in \tau} \frac{\pi(z_{(\dot{\tau}, \ddot{\tau})}, v_{\ddot{\tau}} | v_{\dot{\tau}}, y_{\dot{\tau}}, \theta^{\dagger})}{\pi(z_{(\dot{\tau}, \ddot{\tau})}, v_{\ddot{\tau}} | v_{\dot{\tau}}, y_{\dot{\tau}}, \theta)},
  \end{split}
\end{equation}
where the immediate way of constructing a 2-coin algorithm would be to factorize as in the hidden data update, such that $\pi(z_{(\dot{\tau}, \ddot{\tau})}, v_{\ddot{\tau}} | v_{\dot{\tau}}, y_{\dot{\tau}}, \theta^{\dagger}) = (d_{\theta} \times q_{\theta})(z_{(\dot{\tau}, \ddot{\tau})}, y_{\dot{\tau}}, v_{\braces{\dot{\tau}, \ddot{\tau}}})$. This solution has two major downsides, the first of which is that the integrand in
\begin{equation}
  \begin{aligned}
  \prod_{(\dot{\tau} \sim \ddot{\tau}) \in \tau} q_{\theta}(z_{(\dot{\tau}, \ddot{\tau})}, y_{\dot{\tau}}, v_{\braces{\dot{\tau}, \ddot{\tau}}}) = \exp\braces*{\sum_{(\dot{\tau} \sim \ddot{\tau}) \in \tau} \int_{\dot{\tau}}^{\ddot{\tau}} \lb{\tilde{\varphi}}_{\theta}(z_{(\dot{\tau}, \ddot{\tau})}, y_{\dot{\tau}}, v_{\braces{\dot{\tau}, \ddot{\tau}}}) - \tilde{\varphi}_{\theta}(z_{t}, y_{\dot{\tau}}, v_{\braces{\dot{\tau}, \ddot{\tau}}}) \dd{t}}
  \end{aligned}
\end{equation}
accumulates linearly in $\omega$. Therefore, the compound coin has probability of heads of order $e^{-\class{O}(\omega)}$, and following \eqref{eq:2ctime}, the superordinate 2-coin algorithm has expected number of iterations of order $e^{\class{O}(\omega)}$. Secondly, the cost of this 2-coin algorithm does not decrease in the step size $|\theta^{\dagger} - \theta|$. This is due to the fact that the locality of the $\theta$-update is not reflected in the factorization.

A partial solution consists of working with ratios of intractable quantities, as in the \emph{exchange algorithm} of \citep{murray2012mcmc}. If we define
\begin{equation}
  \xi_{\theta^{\dagger}, \theta}(z_{t}, y_{\dot{\tau}}, v_{\braces{\dot{\tau}, \ddot{\tau}}}) = (\tilde{\varphi}_{\theta^{\dagger}} - \tilde{\varphi}_{\theta})(z_{t}, y_{\dot{\tau}}, v_{\braces{\dot{\tau}, \ddot{\tau}}}) \qquad (t \in (\dot{\tau}, \ddot{\tau}))
\end{equation}
we obtain the valid 2-coin factorization
\begin{equation}
  \begin{split}
    \frac{\alpha_{\Theta}(\theta, \theta^{\dagger})}{\alpha_{\Theta}(\theta^{\dagger}, \theta)}
    & = \frac{\kappa(\theta | \theta^{\dagger})}{\kappa(\theta^{\dagger} | \theta)} \frac{\pi(\theta^{\dagger})}{\pi(\theta)} \prod_{(\dot{\tau} \sim \ddot{\tau}) \in \tau} \frac{h_{\theta^{\dagger}}(y_{\dot{\tau}}, v_{\braces{\dot{\tau}, \ddot{\tau}}})}{h_{\theta}(y_{\dot{\tau}}, v_{\braces{\dot{\tau}, \ddot{\tau}}})} \exp\braces*{-\int_{\dot{\tau}}^{\ddot{\tau}} \xi_{\theta^{\dagger}, \theta}(z_{t}, y_{\dot{\tau}}, v_{\braces{\dot{\tau}, \ddot{\tau}}}) \dd{t}} \\
    & = \prod_{(\dot{\tau} \sim \ddot{\tau}) \in \tau} \frac{\overbrace{\braces{\kappa(\theta | \theta^{\dagger}) \pi(\theta^{\dagger})}^{1/(|\tau| - 1)} h_{\theta^{\dagger}}(y_{\dot{\tau}}, v_{\braces{\dot{\tau}, \ddot{\tau}}})}^{c_{1}^{(\dot{\tau}, \ddot{\tau})}}}{\underbrace{\braces{\kappa(\theta^{\dagger} | \theta) \pi(\theta)}^{1/(|\tau| - 1)} h_{\theta}(y_{\dot{\tau}}, v_{\braces{\dot{\tau}, \ddot{\tau}}})}_{c_{2}^{(\dot{\tau}, \ddot{\tau})}}} \frac{\overbrace{e^{-\int_{\dot{\tau}}^{\ddot{\tau}} \xi_{\theta^{\dagger}, \theta}(z_{t}, y_{\dot{\tau}}, v_{\braces{\dot{\tau}, \ddot{\tau}}})^{(+)} \dd{t}}}^{p_{1}^{(\dot{\tau}, \ddot{\tau})}}}{\underbrace{e^{-\int_{\dot{\tau}}^{\ddot{\tau}} \xi_{\theta, \theta^{\dagger}}(z_{t}, y_{\dot{\tau}}, v_{\braces{\dot{\tau}, \ddot{\tau}}})^{(+)} \dd{t}}}_{p_{2}^{(\dot{\tau}, \ddot{\tau})}}}, 
  \end{split}
\end{equation}
where the superscript $(+)$ denotes the positive part. Loose bounds on the integrands in the range $t \in (\dot{\tau}, \ddot{\tau})$ are given by
\begin{gather}
  \xi_{\theta^{\dagger}, \theta}(z_{t}, y_{\dot{\tau}}, v_{\braces{\dot{\tau}, \ddot{\tau}}}) \leq (\ub{\tilde{\varphi}}_{\theta^{\dagger}} - \lb{\tilde{\varphi}}_{\theta})(z_{(\dot{\tau}, \ddot{\tau})}, y_{\dot{\tau}}, v_{\braces{\dot{\tau}, \ddot{\tau}}}).
\end{gather}
The following proposition illustrates the scaling advantage of this factorization.

\begin{proposition}
  Let $\pi(\theta)$ be supported on a compact set $\class{T} \subset \reals$ with $|\partial_{\theta} \tilde{\varphi}_{\theta}(z_{t}, y_{\dot{\tau}}, v_{\braces{\dot{\tau}, \ddot{\tau}}})|$ uniformly bounded, fix the observation interval $\ddot{s} - \dot{s}$ for all $(\ddot{s} \sim \dot{s}) \in s$, and set the proposal $\theta^{\dagger} | \theta \sim \op{Unif}{\theta \pm \varsigma / \sqrt{\omega}}$. Then, $\int_{0}^{\omega} |\xi_{\theta^{\dagger}, \theta}(z_{t}, y_{\dot{\tau}}, v_{\braces{\dot{\tau}, \ddot{\tau}}})| \dd{t} = \class{O}(\sqrt{\omega})$.
  \begin{proof}
    By the mean value theorem,
    \begin{equation}
      |\xi_{\theta^{\dagger}, \theta}(z_{t}, y_{\dot{\tau}}, v_{\braces{\dot{\tau}, \ddot{\tau}}})| \leq |\theta^{\dagger} - \theta| \sup_{\theta \in \class{T}} \bars*{\partial_{\theta} \tilde{\varphi}_{\theta}(z_{t}, y_{\dot{\tau}}, v_{\braces{\dot{\tau}, \ddot{\tau}}})}, \qquad (t \in (\dot{\tau}, \ddot{\tau}))
    \end{equation}
    and since $\theta^{\dagger} - \theta \sim \op{Unif}{0, \varsigma / \sqrt{\omega}}$, $|\theta^{\dagger} - \theta| \leq \varsigma / \sqrt{\omega}$. By the uniform bound on the gradient, the claim follows.
  \end{proof}
\end{proposition}

Scaling the step size as $\class{O}(\omega^{-1/2})$ is justified by the typical Bernstein-von Mises posterior concentration rate of $\omega^{-1/2}$ when appending observations at constant intervals, and usually results in an acceptance rate of constant order as $\omega \to \infty$ \citep{schmon2022optimal}. The bounded gradient assumption is more stringent, but a similar result could be shown to hold on average, and it illustrates that the alternative factorization exploits posterior concentration. Such a factorization could also be adopted in the hidden data update - this would be particularly useful if that update is highly localized, such that the integrands contributed by state and proposal cancel out. We further improve the scaling of the parameter update by applying the divide-and-conquer Bernoulli factory of \citep{stumpf2025scalable}, which takes as inputs the full collection of weights $\braces{c_{1}^{(\dot{\tau}, \ddot{\tau})}}$ and $\braces{c_{2}^{(\dot{\tau}, \ddot{\tau})}}$ as well as coins $\braces{p_{1}^{(\dot{\tau}, \ddot{\tau})}}$ and $\braces{p_{2}^{(\dot{\tau}, \ddot{\tau})}}$, and constructs the acceptance coin hierarchically from smaller batches. In \citep{stumpf2025scalable}, it was observed that the combination of those remedies improves the cost of the $\theta$-update to as low as $\class{O}(\omega)$ for ordinary diffusion models, which we expect to hold for switching models more broadly, as opposed to a scaling of $e^{\class{O}(\sqrt{\omega})}$  for the naive 2-coin implementation.

\subsection{Generator Update}
\label{ssec:gen}

The update to $\pi(\lambda | y) \propto \pi(y | \lambda) \pi(\lambda)$ does not involve the augmented transition densities, it is therefore the easiest update to carry out. For jump times $r$, the density $\pi(y | \lambda)$ with respect to $\meas{L}$ is given by
\begin{equation}
  \pi(y | \lambda) = \exp\braces*{\int_{0}^{\omega} (1 - \lambda_{y_{t}}) \dd{t}} \prod_{(\dot{r} \sim \ddot{r}) \in r} \lambda_{y_{\dot{r}} y_{\ddot{r}}}.
\end{equation}
Defining the cumulative holding times $\chi_{i} = \int_{0}^{\omega} \op{1}[\braces{i}](y_{t}) \dd{t}$ and the jump counts $n_{ij}$ from regime $i$ to $j$, we find that they are a sufficient statistic:
\begin{gather}
  \pi(y | \lambda) \propto \prod_{i} \parens*{e^{-\lambda_{i} \chi_{i}} \prod_{j \neq i} \lambda_{ij}^{n_{ij}}}.
\end{gather}
We set the conjugate product prior
\begin{equation}
  \pi(\lambda) = \prod_{i \neq j} \op{Gamma}{\lambda_{ij}; \alpha, \beta},
\end{equation}
such that the free elements of $\Lambda$ are independent a priori. The posterior distribution then becomes
\begin{equation}
  \pi(\lambda | y) = \prod_{i \neq j} \op{Gamma}{\lambda_{ij}; n_{ij} + \alpha, \chi_{i} + \beta}.
\end{equation}
The reader interested in applications should note that this prior is necessarily informative - $Y$ is usually ill-identified by the data alone. This is especially the case if $\lambda$ allows for regimes that are ephemeral relative to the observation frequency on the diffusion path and therefore vacuous. Accordingly, the prior expectation of $\Lambda_{i}$, given by
\begin{equation}
  \op{E}{\Lambda_{i}} = \sum_{i \neq j} \op{E}{\Lambda_{ij}} = \sum_{i \neq j} \frac{\alpha_{ij}}{\beta_{ij}},
\end{equation}
should be chosen such that it is smaller than the mean observation rate.

\subsection{Practical Considerations}
\label{ssec:pracmc}

In diffusion settings, the main practical challenge that Bernoulli factory MCMC faces is \emph{proposal sensitivity} - the notion that the 2-coin algorithm runtime can explode when proposing a move to some regions of the state space, especially those associated with large diffusion drift. This is due to the fact that in such regions, the discrepancy between the diffusion measure and the Brownian bridge proposals is large, and the success probability of Poisson coins is low. Under a correctly specified model, the diffusion path avoids regions of $\class{V}$ with high drift, and conversely, parameters associated with large drift usually have low posterior probability. Accordingly, the $\theta$-proposals in particular need to be carefully chosen. We take two steps to mitigate this issue, the first of which is to apply the \emph{Portkey Barker algorithm} of \citep{vats2020efficient}, which truncates the infinite 2-coin loop at the cost of lesser MCMC efficiency. In practice, the main advantage of this modification is that it discards proposals with potentially very low acceptance probability more quickly. Moreover, the Portkey approach can be extended to the divide-and-conquer Bernoulli factory. The other measure consists of initializing the chain and pre-adapting the proposals by running a chain on an Euler-approximated posterior first. We specify the approximation in Supplement \ref{sec:approx}. This allows us to then start the exact chain at a better location in the state space, with reasonable tuning parameters. We adopt both mitigation strategies in Sections \ref{sec:demo} and \ref{sec:simstud}.

More specific challenges apply to Markov switching diffusions, which can exhibit large drift even under the correct specification. This occurs e.g. when switching between two regimes with different stationary means, upon which the process may experience strong drift towards the new equilibrium. This is partially mitigated by refining the bounds on the diffusion path, as recommended by \citep{gonccalves2023exact}, which increases the success probability of Poisson coins. The shortcoming of this intervention is the difficulty of tuning it adaptively during the MCMC run, and usually the chain has to be fully reset to modify the extent of the refinement. We also note that fairly strong posterior dependence between $\Theta$ and $Y$ can arise, which negatively affects mixing of the Gibbs chain. This tendency, while not catastrophic, is observed in the experiment in Section \ref{sec:demo}. Indeed, the corresponding model is invariant to label permutations, and the Markov chain will typically only visit one of the posterior's equivalent modes. Moreover, it is common for the chain to drop one of the states during the transient phase, upon which the update for the parameters corresponding to that state may walk randomly. Therefore, during adaptation, we set the parameters of any inactive state equal to the parameters of an active state, which results in quick re-introduction of the inactive state.

\section{Exact MAP and Maximum Likelihood Estimation}
\label{sec:map}

A natural companion problem to posterior sampling is \emph{maximum a posteriori} (MAP) estimation, i.e. finding the set of values $(\theta^{\ddagger}, \lambda^{\ddagger})$ such that
\begin{equation}
  (\theta^{\ddagger}, \lambda^{\ddagger}) = \op*{argmax}[\theta, \lambda] \ \pi(\theta, \lambda, v_{s \setdiff \braces{0}} | v_{0}).
\end{equation}
The MAP estimator also corresponds to the maximum likelihood estimator when setting $\pi(\theta, \lambda) \propto 1$. In this section, we adapt an approach originally proposed for Ito diffusions in \citep{beskos2006exact}. It consists of constructing a Monte Carlo EM algorithm, which alternates between an approximate E-step, where we construct a Monte Carlo estimator $\est{Q}$ of the function
\begin{equation}
  Q(\theta^{\dagger}, \lambda^{\dagger}, \theta, \lambda) = \op{E}[V_{R}, Z, Y]{\log \pi(V_{R}, Z, Y, \theta^{\dagger}, \lambda^{\dagger}, v_{s \setdiff \braces{0}} | v_{0}) | v_{s}, \theta, \lambda},
\end{equation}
and an M-step, where we solve for
\begin{equation}
  \op*{argmax}[\theta^{\dagger}, \lambda^{\dagger}] \ \est{Q}(\theta^{\dagger}, \lambda^{\dagger}, \theta, \lambda).
\end{equation}
One benefit of the MCEM approach is that there is a substantial methodological overlap with posterior sampling, and we can make use of the hidden data update of Section \ref{ssec:hidden} in devising the E-step. The MCEM algorithm also exploits the same conditional independence structure. Moreover, because it merely requires an unbiased estimate of the log-likelihood, the MCEM algorithm avoids some of the scaling issues with the MCMC parameter update, discussed in Section \ref{ssec:param}. Conversely, even an ``exact'' MCEM algorithm is not guaranteed to converge to the global maximum, though convergence to a local maximum can be proven in some circumstances \citep{fort2003convergence}.

We discuss the construction of $\est{Q}$ in Section \ref{ssec:estep} before proceeding with its maximization in Section \ref{ssec:mstep}, and close out with a discussion of various implementation aspects in Section \ref{ssec:pracem}.

\subsection{E-Step}
\label{ssec:estep}

In standard EM algorithms, the E-step consists of constructing a lower bound on the objective $\pi(\theta, \lambda, v_{s \setdiff \braces{0}} | v_{0})$. It is obtained by averaging the joint density over the posterior of the latent variables, i.e.
\begin{equation}
  \begin{split}
    Q(\theta^{\dagger}, \lambda^{\dagger}, \theta, \lambda)
    & = \op{E}[V_{R}, Z, Y]{\log \pi(V_{R}, Z, Y, \theta^{\dagger}, \lambda^{\dagger}, v_{s \setdiff \braces{0}} | v_{0}) | v_{s}, \theta, \lambda} \\
    & = \op{E}[V_{R}, Z, Y]{\log \pi(V_{R}, Z, v_{s \setdiff \braces{0}} | Y, \theta^{\dagger}, v_{0}) + \log \pi(Y | \lambda^{\dagger}) | v_{s}, \theta, \lambda} \\
    & \qquad + \log \pi(\theta^{\dagger}) + \log \pi(\lambda^{\dagger}).
  \end{split}
\end{equation}
where we take expectations with respect to $\pi(v_{r}, z, y | v_{s}, \theta, \lambda)$. Because the $Q$-function decomposes into separate functions of $\theta^{\dagger}$ and $\lambda^{\dagger}$, we may define separate Q-functions
\begin{gather}
  Q_{\Theta}(\theta^{\dagger}, \theta) = \op{E}[V_{R}, Z, Y]{\log \pi(V_{R}, Z, v_{s \setdiff \braces{0}} | v_{0}, y, \theta^{\dagger}) | v_{s}, \theta, \lambda} + \log \pi(\theta^{\dagger}), \\
  Q_{\Lambda}(\lambda^{\dagger}, \lambda) = \op{E}[V_{R}, Z, Y]{\log \pi(Y | \lambda^{\dagger}) | v_{s}, \theta, \lambda} + \log \pi(\lambda^{\dagger}).
\end{gather}
When the expectation is not tractable, MCEM algorithms replace $Q$ with an estimator thereof. In this instance, the expectation is taken with respect to $\pi(v_{r}, z, y | v_{s}, \theta, \lambda)$, for which we developed a sampling algorithm in Section \ref{ssec:hidden}. Therefore, we can simulate a Markov chain with stationary distribution $\pi(v_{r}, z, y | v_{s}, \theta, \lambda)$, generate a sequence $\braces{(v_{r}^{(l)}, z^{(l)}, y^{(l)})}$ of $\ell$ samples, and replace $Q_{\Theta}$ with the ergodic average
\begin{equation}
  \log \pi(\theta^{\dagger}) + \ell^{-1} \sum_{l=1}^{\ell} \log \pi(V_{R}^{\dagger}, Z^{\dagger}, v_{s \setdiff \braces{0}} | v_{0}, y^{\dagger}, \theta^{\dagger}).
\end{equation}
This differs from the Importance sampling approach taken e.g. by \citep{gonccalves2023exact}, though in this instance, where the data can be highly informative about $Y$, we deem MCMC estimation to be safer. Since $\pi(v_{\tau \setdiff \braces{0}}, z | v_{0}, y, \theta)$ is itself intractable, we require a further estimation step. Conveniently, unbiased estimation thereof is easier on the log scale. The log complete transition density is given by
\begin{equation}
  \log \pi(z_{(\dot{\tau}, \ddot{\tau})}, v_{\ddot{\tau}} | v_{\dot{\tau}}, y_{\dot{\tau}}, \theta) = \log h_{\theta}(y_{\dot{\tau}}, v_{\braces{\dot{\tau}, \ddot{\tau}}}) - \int_{\dot{\tau}}^{\ddot{\tau}} \tilde{\varphi}_{\theta}(z_{t}, y_{\dot{\tau}}, v_{\braces{\dot{\tau}, \ddot{\tau}}}) \dd{t},
\end{equation}
where the time integral can be estimated without bias by uniform subsampling along the path:
\begin{equation}
  \int_{\dot{\tau}}^{\ddot{\tau}} \tilde{\varphi}_{\theta}(z_{t}, y_{\dot{\tau}}, v_{\braces{\dot{\tau}, \ddot{\tau}}}) \dd{t} = \op{E}[U]{(\dot{\tau} - \ddot{\tau}) \tilde{\varphi}_{\theta}(z_{U}, y_{\dot{\tau}}, v_{\braces{\dot{\tau}, \ddot{\tau}}})}, \quad U \sim \op{Unif}{\dot{\tau}, \ddot{\tau}}.
\end{equation}
Thus, we define the log augmented transition density estimator
\begin{equation}
  \est{\ell}_{u}(z_{(\dot{\tau}, \ddot{\tau})}, v_{\ddot{\tau}} | v_{\dot{\tau}}, y_{\dot{\tau}}, \theta) = \log h_{\theta}(y_{\dot{\tau}}, v_{\braces{\dot{\tau}, \ddot{\tau}}}) - (\ddot{\tau} - \dot{\tau}) \tilde{\varphi}_{\theta}(z_{u}, y_{\dot{\tau}}, v_{\braces{\dot{\tau}, \ddot{\tau}}}),
\end{equation}
and enrich each Markov chain sample with a sequence $\braces{u_{(\dot{\tau}, \ddot{\tau})}^{(l)}: (\dot{\tau} \sim \ddot{\tau}) \in \tau^{(l)}}$. Notice how the variance of this estimator is easily controlled, and how this differs from an MCEM algorithm that further augments the complete likelihood to obtain a tractable $Q$-estimator, which is likely to be slower to converge due to increased dependency between the E- and M-steps.

Assembling those elements, we obtain the unbiased $Q$-estimators
\begin{gather}
  \est{Q}_{\Theta}(\theta^{\dagger}) = \log \pi(\theta^{\dagger}) + \ell^{-1} \sum_{l=1}^{\ell} \sum_{(\dot{\tau} \sim \ddot{\tau}) \in \tau^{(l)}} \est{\ell}_{u_{(\dot{\tau}, \ddot{\tau})}^{(l)}}(z_{(\dot{\tau}, \ddot{\tau})}^{(l)}, v_{\ddot{\tau}}^{(l)} | v_{\dot{\tau}}^{(l)}, y_{\dot{\tau}}^{(l)}, \theta), \\
  \est{Q}_{\Lambda}(\lambda^{\dagger}) = \log \pi(\lambda^{\dagger}) + \ell^{-1} \sum_{l=1}^{\ell} \log \pi(y^{(l)} | \lambda^{\dagger}).
\end{gather}
Note that once $(\theta^{\dagger}, \lambda^{\dagger})$ has been chosen in the M-step, the chain on $\pi(v_{r}, z, y | v_{s}, \theta^{\dagger}, \lambda^{\dagger})$ is restarted with the new parameter values and from the last $(V_{R}, Z, Y)$-sample to generate the next $Q$-estimate.

\subsection{M-Step}
\label{ssec:mstep}

Having constructed the $\est{Q}$-estimators in the E-step, we proceed to maximizing the estimated lower bound functions by solving the optimization problems
\begin{equation}
  \braces*{\op*{argmax}[\theta^{\dagger}] \est{Q}_{\Theta}(\theta^{\dagger}), \quad \op*{argmax}[\lambda^{\dagger}] \est{Q}_{\Lambda}(\lambda^{\dagger})}.
\end{equation}
If we assume that $\pi(\theta)$, $\mu_{\theta}$, $\sigma_{\theta}$ and $\rho_{\theta}$ are continuous in $\theta$, as is usually the case in applications, $\est{Q}_{\Theta}(\theta^{\dagger})$ is also continuous and amenable to optimization with a numerical routine, e.g. BFGS. Gradients can be obtained numerically during optimization, or even symbolically prior to the MCMC run, based on symbolic specifications of $(\mu_{\theta}, \sigma_{\theta}, \rho_{\theta})$. 

Conversely, we may optimize $\est{Q}_{\Lambda}(\lambda^{\dagger})$ exactly for a range of priors. Recall from Section \ref{ssec:gen} that the complete data likelihood of the realization $y^{(l)}$ may be expressed in terms of the jump counts $n_{ij}$ from regime $i$ to $j$ and the cumulative regime holding times $\chi_{i}$. As before, we set $\lambda_{ij} \sim \op{Gamma}{\alpha, \beta}$ a priori. The corresponding estimator is
\begin{equation}
  \est{Q}_{\Lambda}(\lambda^{\dagger}) = \sum_{i \neq j} \parens*{\ell^{-1} \sum_{l=1}^{\ell} (n_{ij}^{(l)} \log \lambda_{ij}^{\dagger} + \chi_{i}^{(l)} \lambda_{i}^{\dagger}) + (\alpha - 1) \log \lambda_{ij}^{\dagger} - \beta \lambda_{ij}^{\dagger}}.
\end{equation}
Taking derivatives results in independent FOCs and yields the optimal values
\begin{equation}
  \lambda_{ij}^{\dagger} = \begin{dcases} \frac{\alpha - 1 + \ell^{-1} \sum_{l=1}^{\ell} n_{ij}^{(l)}}{\beta + \ell^{-1} \sum_{l=1}^{\ell} \chi_{i}^{(l)}} & \text{if $\alpha - 1 + \ell^{-1} \sum_{l=1}^{\ell} n_{ij}^{(l)} > 0$} \\ 0 & \text{otherwise} \end{dcases}.
\end{equation}
This solution reveals a limitation of the algorithm: If $\lambda_{ij}^{\dagger}$ is a boundary solution for all $i$, then regime $i$ has 0 probability of being visited under $\pi(y | \lambda^{\dagger})$ and it will be ignored in all subsequent iterations, i.e. the algorithm has absorbing states. We may prevent this behavior by choosing a value $\alpha > 1$, but this excludes the pure maximum likelihood case which corresponds to $\alpha = 1$, $\beta = 0$. If pure ML estimation is required, the absorbing states can be avoided by using a \emph{stable} algorithm, in the terminology of \citep{fort2003convergence}. Such an algorithm resets the generator to a safe value when the M-step enters a forbidden, progressively vanishing set.

\subsection{Practical Considerations}
\label{ssec:pracem}

As in the posterior sampling case, we deem it helpful to precede the main optimization run with an Euler-approximated run, as sketched out in Supplement \ref{sec:approx}. Since we do not need the algorithm to fully converge at this stage, $\ell$ may be kept constant there. This will typically initiate the main run in fairly close proximity to the MAP, with good tuning parameters for the hidden data update. Tuning parameters are best kept and further adapted across E-steps. In the main run, $\ell$ has to be increased at each E-step to achieve convergence. At the $m$-th E-step, \citep{fort2003convergence} recommend increasing $\ell_{m}$ at a rate less than exponential, such that $\lim_{m \to \infty} \ell_{m+1} / \ell_{m} = 1$. Where large $\ell$ are required to reach convergence, the $Q$-estimate may be constructed from a thinned chain in order to reduce computational burden in the M-step.

\section{Demonstration: Animal Tracking}
\label{sec:demo}

\begin{figure}
  \centering
  \includegraphics[scale=.66]{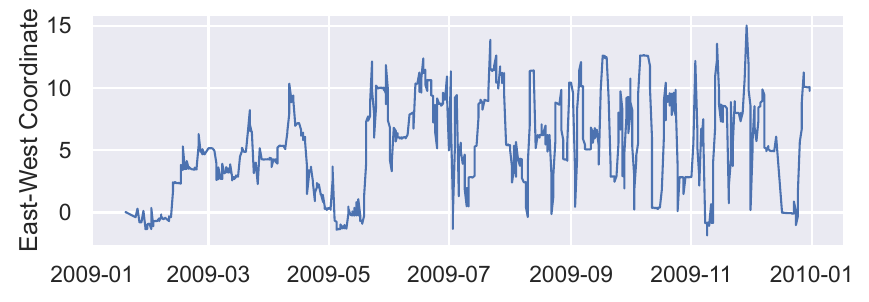}
  \caption{[Mountain Lion tracking model] East-West Movement of Mountain Lion f109 in 2009.}
  \label{fig:f109-data}
\end{figure}


In this section, we demonstrate the MCMC and the MCEM algorithms on a moving-resting model for animal movement. The observations were collected from the tagged mountain lion ``f109'' over multiple years and at irregular intervals, and are included e.g. in the \texttt{smam} package on \texttt{CRAN}. The time series for the lion's east-west location is shown in Figure \ref{fig:f109-data}. Markov switching models have previously been applied e.g. by \citep{yan2014moving} to capture the alternating moving-resting dynamics in the movement. In \citep{yan2014moving}, the model takes the form of a tractable 2-regime model, the first of which is a ``moving'' regime modelled by Brownian motion, and the second is a ``resting'' regime where the position is fixed. We extend this model by allowing for weak mean reversion, reflecting the fact that the process does not appear to be transient, and we leave all regimes fully symmetrical a priori. Hence, the distinction between ``moving'' and ``resting'' states is an entirely implicit property of the posterior, rather than being imposed a priori. The model's SDE specification is
\begin{equation}
  \dd{V_{t}} = \rho_{Y_{t}} (\beta_{Y_{t}} \op{tanh}{\mu_{Y_{t}} - V_{t}} \dd{t} + \dd{W_{t}}), \qquad (\beta_{i}, \rho_{i} > 0, \quad i = 0, 1)
\end{equation}
where the drift function is bounded in absolute value by $\rho_{Y_{t}} \beta_{Y_{t}}$, and time is indexed in hours. For each of the regimes, the diffusion is driven by a separate set of parameters $\theta_{i} = \braces{\rho_{i}, \beta_{i}, \mu_{i}}$. For a given regime, the stationary density is $\pi(a) \propto \op{cosh}[][-2\beta_{i}/\rho_{i}]{a - \mu_{i}}$, as obtained by finding the stationary solution of the Fokker-Planck equation, which is similar to a Laplacian density. The transition density for this model is intractable, it therefore falls within the scope of our method. We use symmetrical priors for all regimes:
\begin{equation}
  \mu_{i}, \log \beta_{i}, \log \rho_{i} \sim \op{N}{0, 1}, \quad \lambda_{i} \sim \op{Exp}{48}, \qquad (i = 0, 1)
\end{equation}
This implies a prior expectation of one regime transition every $48$ hours. The specification results in a posterior that is invariant to label permutations and therefore multimodal. Nonetheless, when the modes are sufficiently separated, the algorithm typically doesn't permute the labels.

Due to prior independence of the regime parameters and the constant Lamperti transform, we benefit from the additional conditional independence
\begin{gather}
  \pi(\theta | v_{\tau}, z, y) = \prod_{i=1}^{k} \pi(\theta_{i} | v_{\tau}, z, y).
\end{gather}
Therefore, given independent proposals $\kappa(\theta_{i}^{\dagger} | \theta_{i})$, we can carry out independent parameter updates for each of the regimes. Similar simplifications occur in the M-step of the MCEM algorithm.

We do not provide further details on the model-specific form of the functions ($\eta_{\theta}$, $\varphi_{\theta}$, ...) since these are constructed automatically by the implementation, merely requiring the specification $\theta$, $\mu_{\theta}$, $\sigma$ and $\rho_{\theta}$. This is provided with a few lines of symbolic code, similar to the following Python snippet:
\begin{lstlisting}
  v, x = sympy.symbols('v x', real=True)
  b, r = sympy.symbols('b r', positive=True)
  m = sympy.symbols('m', real=True)
  thi = sympy.Array([m, b, r])
  mu = r * b * sympy.tanh(m - v)
  sig = sympy.Integer(1)
  rho = r
\end{lstlisting}
The resulting functions are then provided to a model-agnostic backend.

\subsection{MCMC Results}
\label{ssec:demomc}

\begin{figure}
  \centering
  \includegraphics[scale=.66]{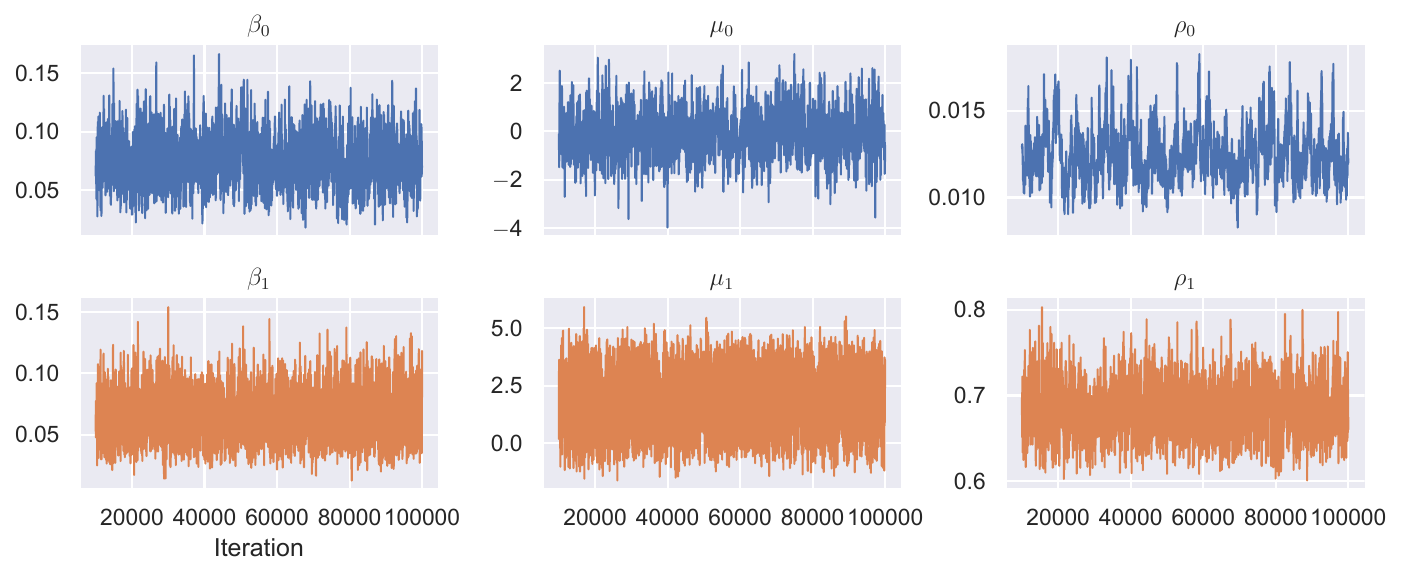}
  \caption{[Mountain Lion tracking model] Parameter traces vs. MCMC iteration.}
  \label{fig:f109-mcmc-traces}
\end{figure}

We run the MCMC algorithm for 100000 iterations, after 10000 iterations of pre-adaptation with the approximate algrorithm. We target an acceptance probability of .2 and apply a ``portkey'' setting of .001, and carry out 4 splits in the divide-and-conquer 2-coin algorithm when updating $\theta$.

The algorithm converges on a solution with a low volatility ``resting'' regime, and a high volatility ``movement'' regime, as presumed by the model of \citep{yan2014moving}. Figure \ref{fig:f109-mcmc-latent} shows how the resting regime is favored in periods of little observed movement, and vice versa, and how the predictive distribution of $V_{t}$ naturally adjusts. Within those regimes, we observe adequate mixing of $\Theta$, as seen in Figure \ref{fig:f109-mcmc-traces}. Since the marginals of $\beta_{0}, \beta_{1}$ and $\kappa_{0}, \kappa_{1}$ overlap, the respective latent states are most strongly identified by $\rho_{0}, \rho_{1}$, resulting in stronger posterior dependence on $Y$ and slower mixing of the volatility parameters.

\subsection{MCEM Results}
\label{ssec:demoem}

\begin{figure}
  \centering
  \includegraphics[scale=.66]{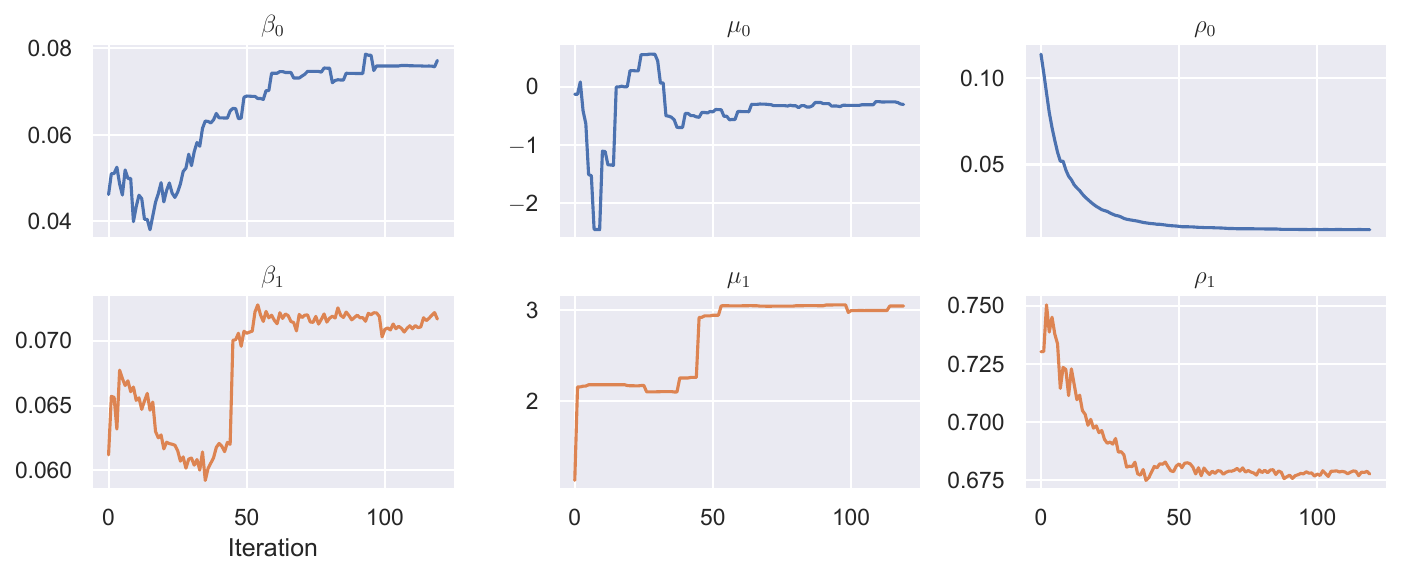}
  \caption{[Mountain Lion tracking model] Parameter estimate vs. MCEM iteration.}
  \label{fig:f109-mcem-params}
\end{figure}

We run the MCEM algorithm for 100 iterations, after 100 iterations of pre-adaptation with the approximate algorithm. The number of iterations $\ell_{m}$ in the $m$-th E-step is $m^{1.25}$ in the main run and 1 in the approximate run. In the MCMC chain within the E-step, we again target an acceptance probability of .2 and apply a ``portkey'' setting of .001. In the M-step, we optimize the ELBO estimate using the BFGS algorithm with numerical gradients.

The algorithm converges on similar ``moving'' and ``resting'' regimes as in the posterior sampling case. The estimate of the evidence lower bound (ELBO) stabilizes towards the end of the run, as do the parameter estimates shown in Figure \ref{fig:f109-mcem-params}. The parameter estimates coincide with the location of the modes in Figure \ref{fig:f109-bias}, as estimated from the MCMC output.

\subsection{Approximation Bias}
\label{ssec:bias}

\begin{figure}
  \centering
  \includegraphics[scale=.66]{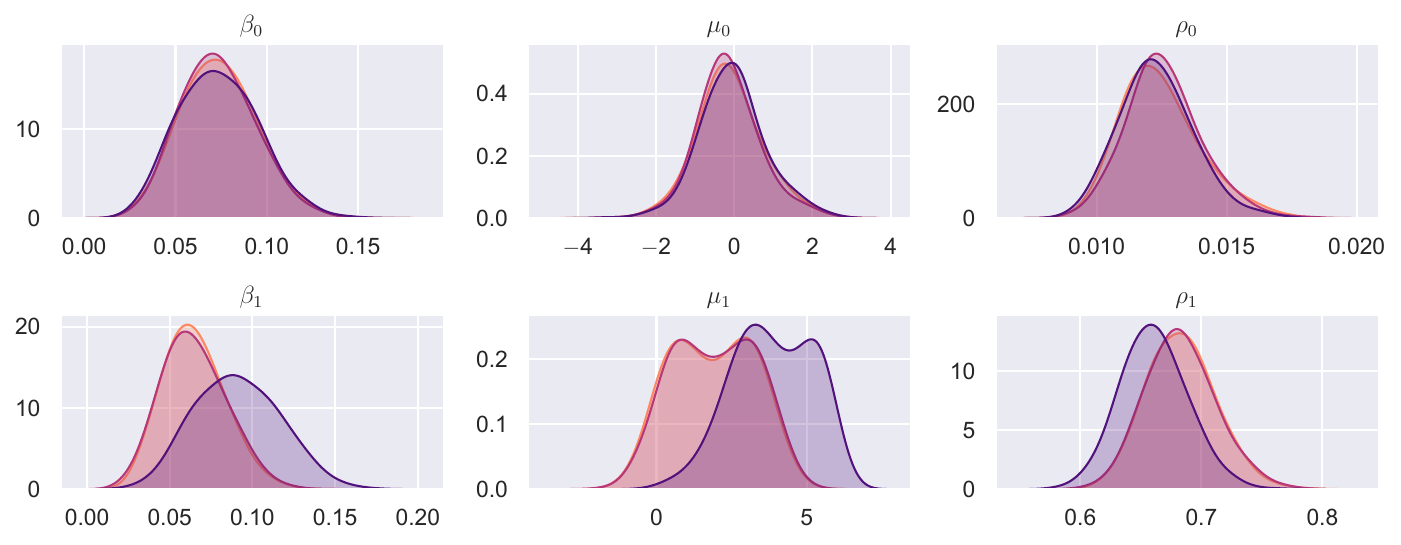}
  \caption{[Mountain Lion tracking model] Posterior marginals of $\theta$ for the exact algorithm (bright), the approximate algorithm without imputation (dark), and the approximate algorithm that imputes observations at rate 4 per unit time (intermediate).}
  \label{fig:f109-bias}
\end{figure}

In order to examine the practical impact of approximation biases in Euler-approximated algorithms, we also devised an approximate MCMC algorithm which we outline in Supplement \ref{sec:approx}. This algorithm imputes additional observations at a given rate, with higher imputation rates raising computational cost, but giving a closer approximation to the exact posterior. In the limit of higher imputation rates, the algorithm targets the exact posterior. Indeed, under the hood, the exact algorithm imputes stochastically rather than deterministically, with a rate that tends to increase with Euler approximation bias.

In Figure \ref{fig:f109-bias} we compare the posterior marginals of $\theta$ as obtained by the exact algorithm as well as approximate algorithms with different imputation rates. We run the approximate algorithms for 50000 iterations, a target acceptance probability of .234 and with imputation rates 0 and 4 per unit time, respectively. The figure indicates that in this instance, doing no imputation results in a sizable bias, while an imputation rate of 4 per unit time is sufficient to eliminate most of that bias. In our implementation, the approximate run with imputation rate 4 runs about 5 times as fast as the exact run. It also has a slightly higher statistical efficiency since it uses the more efficient Metropolis-Hastings, rather than Portkey Barker. We expect that an optimized implementation of the EA3-algorithm would close that performance gap. Moreover, in the absence of the exact benchmark, further imputation would be required to ensure that most of the bias has been eliminated.

\section{Simulation Study}
\label{sec:simstud}

\begin{figure}
  \centering
  \includegraphics[scale=.66]{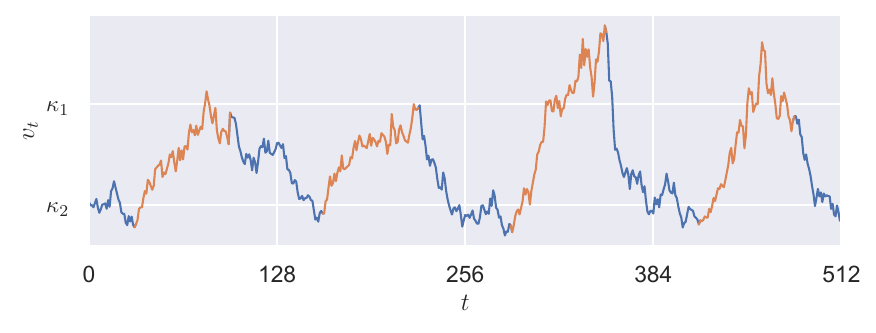}
  \caption{[Simulation Study] Diffusion trajectory for the ``base'' design of the simulation study. Latent regimes are colored blue and orange, respectively.}
  \label{fig:simstud-data}
\end{figure}

In this section, we explore the scaling behavior of the MCMC algorithm in the \emph{outfill regime}, where we append further data to the time series, and the \emph{infill regime}, where we increase observation frequency. The input data simulation protocol uses a deterministic trajectory of $y$ which switches regimes every 64 time units, ensuring that as data is added, it is taken in equal parts during the activity of each regime. Figure \ref{fig:simstud-data} illustrates the deterministic regime switching pattern of the input data for the ``base'' design. We then investigate the efficiency of the marginal and the auxiliary algorithm under both regimes. In the outfill design, the ``base'' dataset is extended by appending 3 more cycles. In the ``infill'' design, 3 additional observations are inserted in between any 2 ``base'' observations.

We adopt a regime-switching version of the \emph{logistic growth model}, defined by the SDE
\begin{equation}
  \dd{V_{t}} = \rho_{Y_{t}} V_{t} (\beta_{Y_{t}} (1 - V_{t} / \kappa_{Y_{t}}) \dd{t} + \dd{W_{t}}), \qquad (\rho_{i}, \beta_{i}, \kappa_{i} > 0, \quad i = 0, 1)
\end{equation}
where $\rho$ is a scale parameter, $\beta$ is the reproduction rate and $\kappa$ is the \emph{carrying capacity} of the environment. For $2\beta_{i} / \rho_{i} > 1$, the corresponding stationary distribution for a fixed regime is $\op{Gamma}{2\beta_{i} / \rho_{i} - 1, 2\beta_{i} / (\rho_{i} \kappa_{i})}$. We also set symmetrical priors for all regimes:
\begin{equation}
  \log \beta_{i}, \log \kappa_{i}, \log \rho_{i} \sim \op{N}{0, 1}, \quad \lambda_{ij} \sim \op{Exp}{2^{6}}, \qquad (i, j = 0, 1).
\end{equation}
Hence, the same factorization over parameter sets $\theta_{i} = \braces{\rho_{i}, \beta_{i}, \kappa_{i}}$ arises as in Section \ref{sec:demo}. We follow the efficiency notion of average CPU seconds per effective sample (S/ES), and estimate it from the output of the MCMC algorithm. Both the average seconds per iteration (S/I) and the average number of iterations per effective sample (I/ES) are estimated from MCMC output for various statistics. The computational cost is part deterministic and part random, with either part affected differently in the scaling regimes. The deterministic part of the cost per iteration is linear in the number of observations in both regimes. For the outfill regime, the optimistic scenario is that random costs remain linear in expectation, while the effective sample size remains constant. For the infill regime, we note that random costs depend on the length of the time series and the uncertainty about the diffusion bridges. Since the length of the series is constant but uncertainty is reduced, random costs should decrease. Conclusions from those experiments have limited external validity, and should be seen as setting a benchmark for the behavior of the algorithms under favorable circumstances, i.e. for models that are fairly smooth in $\theta$ and exhibit sufficient posterior concentration rates. We use the integrated autocorrelation time estimator for effective sample size estimation, as seen e.g. in \citep{sokal1997monte}.

\begin{figure}
  \centering
  \includegraphics[scale=.66]{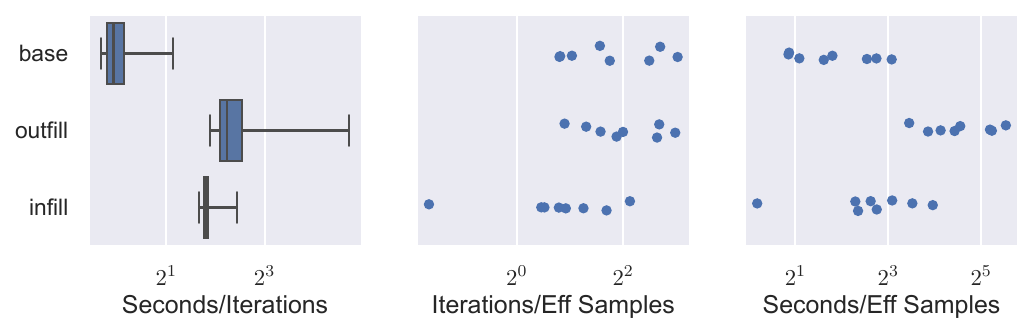}
  \caption{[Simulation Study] Performance measures for the three designs. The distribution of (S/I) over the MCMC run is shown in the left panel for each design. In the middle panel, (I/ES) is shown as a dot for each element of $(\theta, \lambda)$. In the right panel, (S/ES) is shown as a dot for each element of $(\theta, \lambda)$.}
  \label{fig:simstud-performance}
\end{figure}

We obtain estimates from MCMC runs of 10000 iterations for each regime, where we've discarded 1000 iterations for adaptation, and preceded the main run by 1000 iterations of the approximate algorithm. The chain is thinned to every 10th sample to speed up summary computation. We target an acceptance probability of .2, and apply a ``portkey'' setting of $\omega^{-1}$ in the parameter update and .01 in the hidden data update. We use one divide-and-conquer split for the ``base'' and ``infill'' designs, and two splits for the ``outfill'' design, which follows the $\class{O}(\log_{4} \omega)$ scaling recommended in \citep{stumpf2025scalable}.

We show performance metrics in Figure \ref{fig:simstud-performance}. The (S/I) measurements show a long tail in iteration time, though we avoid a substantial increase in outliers in the ``outfill'' regime. The (I/ES) measurements indicate that we meet our objective of achieving stable mixing across the designs. In the outfill regime, the (S/ES) measurements are in line with the $n \log n$ cost scaling observed by \citep{stumpf2025scalable}. In the infill regime (S/ES) is sublinear n $|s| / \omega$ due to tigher bounds on the latent diffusion process.

\section{Discussion}
\label{sec:disc}

We have described exact algorithms for posterior sampling and point estimation in discretely observed Markov switching diffusions, and carried out experiments that demonstrate the computational viability of those methods, especially if progress is made in optimizing low-level components. We have operated in a Bernoulli MCMC framework, which affords advantages in terms of transparency and robustness, and we have addressed its previous limitations in terms of scalability in the length of the time series. The methods were formulated such that no extensive hand tuning or further implementation effort is required, thereby making them more accessible to end users and researchers. We have also found that the exact method can avoid substantial biases in approximate methods.

In spite of those advances, we have restricted our discussion and experiments to univariate diffusions, even as applications often call for multivariate models. While the methodology may be in principle extended to cover some of those models, the conditions that allow for a Lamperti transform are more stringent in multivariate models, as pointed out in \citep{beskos2008factorisation}. Moreover, the Brownian bridge proposals considered herein work best on diffusions for which the Lamperti-transformed process $X$ has state space $\class{X} = \reals$. Covering those cases would further exacerbate the already significant computational and methodological commitment that is required in the more restrictive setting of this paper. 

Furthermore, the methodological breadth and generality of our algorithm comes at a cost in theoretical tractability, which we consider beyond the scope of this paper. Since many of the methods involved do not have uniform characteristics over unbounded parameter spaces, we believe that an analysis of MCMC convergence rates would require strong restrictions on the model class or the prior. Nonetheless, there is theoretical support for some components of the algorithm, such as for Gibbs sampling of $(Z, \Theta)$ in ordinary Ito diffusions, for which \citep{roberts2001inference} found that irreducibility of the Gibbs sampler is synonymous with ``good mixing". Furthermore, for fixed $\theta$, modified versions of our independence sampler for the hidden data can be shown to be uniformly ergodic for various model classes through an argument along the lines of \cite{mengersen1996rates}. This usually requires a more heavy-tailed proposal for $V_{R}$ (diffusion value at jump times) to dominate unbounded terms in Equation \eqref{eq:tractable}, for instance from a Cauchy process. In particular, uniform ergodicity can be shown for models with bounded drift, such as the Tanh diffusion in Section \ref{sec:demo}, and Ornstein-Uhlenbeck type models with up to linear drift. Conversely, models with exponential drift such as the switching logistic growth diffusion from Section \ref{sec:simstud} pose greater challenges.

As for more immediate avenues of research, we note that $\lambda$ can be constructed to give rise to various behaviors - for example, such that $Y$ approximates a continuous-state process \citep{kushner1990numerical, chourdakis2002continuous}. This is naturally accommodated by our methods, and we are currently exploring some of those variants. Another possibility is to address the case of semi-Markov switching diffusions, where the latent process has non-exponential holding times. Most of our algorithm could be applied out the box, though modifications would have to be made to the localized hidden data update described in Supplement \ref{sec:localize}, since bridge simulation of the regime process is more complicated, and different sections of the trajectory cannot be updated independently.

\section*{Acknowledgements}

The authors are grateful to Flávio Gonçalves for his comments on the manuscript. Timothée Stumpf-Fétizon has received funding from the ERC grant PrSc-HDBayLe (No. 101076564). Krzysztof Łatuszynski has been supported by the Royal Society through the Royal Society University Research Fellowship. Gareth O. Roberts has been supported by the UKRI grant OCEAN (EP/Y014650/1) and EPSRC grants Bayes for Health (R018561), CoSInES (R034710), PINCODE (EP/X028119/1), ProbAI (EP/Y028783/1) and EP/V009478/1.
\printbibliography
\appendix
\section{Proof of Proposition \ref{prop:aug}}
\label{sec:proofaug}

We define
\begin{equation}
  \eta_{\theta}(a) = \int_{v^{*}}^{a} \frac{\dd{b}}{\sigma_{\theta}(b)}, \qquad (v^{*} \in \class{V})
\end{equation}
which by Ito's formula yields the reduced process $X = \eta_{\theta}(V)$ with SDE
\begin{gather}
  \dd{X_{t}} = \delta_{\theta}(X_{t}, y_{\dot{\tau}}) \dd{t} + \rho_{\theta}(y_{\dot{\tau}}) \dd{W_{t}}, \qquad (X_{0} = \eta_{\theta}(v_{0}), \; t \in [\dot{\tau}, \ddot{\tau})) \\
  \delta_{\theta}(a, b) = \parens*{\frac{\mu_{\theta}(\cdot, b)}{\sigma_{\theta}} - \frac{\sigma_{\theta}'}{2}} \circ \eta_{\theta}^{-1}(a).
\end{gather}
Notice that while $\delta_{\theta}(X_{t}, y_{t})$ is discontinuous at times $r$, $X$ itself remains continuous. Let $\meas{X}|(x_{\dot{\tau}}, y_{\dot{\tau}}, \theta)$ be induced by $X_{(\dot{\tau}, \ddot{\tau}]}$ for $X_{\dot{\tau}} = x_{\dot{\tau}}$ and $Y_{\dot{\tau}} = y_{\dot{\tau}}$. Furthermore, let $\meas{M}|(x_{\dot{\tau}}, y_{\dot{\tau}}, \theta)$ be the driftless measure induced by $\dd{X_{t}} = \rho_{\theta}(y_{\dot{\tau}}) \dd{W_{t}}$. Having presupposed that the Novikov condition holds, $\meas{M}|(x_{\dot{\tau}}, y_{\dot{\tau}}, \theta) \gg \meas{X}|(x_{\dot{\tau}}, y_{\dot{\tau}}, \theta)$ by the Girsanov Theorem, and the RND between the two measures is
\begin{equation}
  \begin{aligned}
  \der{\meas{X}|(x_{\dot{\tau}}, y_{\dot{\tau}}, \theta)}{\meas{M}|(x_{\dot{\tau}}, y_{\dot{\tau}}, \theta)}(x_{(\dot{\tau}, \ddot{\tau}]})
  & = \exp\braces*{\int_{\dot{\tau}}^{\ddot{\tau}} \frac{\delta_{\theta}(x_{t}, y_{\dot{\tau}})}{\rho_{\theta}(y_{\dot{\tau}})} \dd{W_{t}} + \frac{1}{2} \int_{\dot{\tau}}^{\ddot{\tau}} \frac{\delta_{\theta}^{2}(x_{t}, y_{\dot{\tau}})}{\rho_{\theta}^{2}(y_{\dot{\tau}})} \dd{t}} \\
  & = \exp\braces*{\int_{\dot{\tau}}^{\ddot{\tau}} \frac{\delta_{\theta}(x_{t}, y_{\dot{\tau}})}{\rho_{\theta}^{2}(y_{\dot{\tau}})} \dd{X_{t}} - \frac{1}{2} \int_{\dot{\tau}}^{\ddot{\tau}} \frac{\delta_{\theta}^{2}(x_{t}, y_{\dot{\tau}})}{\rho_{\theta}^{2}(y_{\dot{\tau}})} \dd{t}}.
  \end{aligned}
\end{equation}
We now proceed to eliminating the stochastic integral in the RND. Define the drift antiderivative
\begin{equation}
  \Delta_{\theta}(a, b) = \int_{x^{*}}^{a} \delta_{\theta}(c, b) \dd{c}, \qquad (x^{*} \in \class{X})
\end{equation}
and notice that by Ito's formula,
\begin{equation}
  \frac{\Delta_{\theta}(x_{\ddot{\tau}}, y_{\dot{\tau}}) - \Delta_{\theta}(x_{\dot{\tau}}, y_{\dot{\tau}})}{\rho_{\theta}^{2}(y_{\dot{\tau}})} = \int_{\dot{\tau}}^{\ddot{\tau}} \frac{\delta_{\theta}(x_{t}, y_{\dot{\tau}})}{\rho_{\theta}^{2}(y_{\dot{\tau}})} \dd{X_{t}} + \frac{1}{2} \int_{\dot{\tau}}^{\ddot{\tau}} \partial_{x_{t}} \delta_{\theta}(x_{t}, y_{\dot{\tau}}) \dd{t}.
\end{equation}
Substituting that expression back into the RND, we find its simplified form:
\begin{gather}
  \der{\meas{X}|(x_{\dot{\tau}}, y_{\dot{\tau}}, \theta)}{\meas{M}|(x_{\dot{\tau}}, y_{\dot{\tau}}, \theta)}(x_{(\dot{\tau}, \ddot{\tau}]}) 
  = \exp\braces*{\frac{\Delta_{\theta}(x_{\ddot{\tau}}, y_{\dot{\tau}}) - \Delta_{\theta}(x_{\dot{\tau}}, y_{\dot{\tau}})}{\rho_{\theta}^{2}(y_{\dot{\tau}})} - \int_{\dot{\tau}}^{\ddot{\tau}} \varphi_{\theta}(x_{t}, y_{\dot{\tau}}) \dd{t}}, \\
  \varphi_{\theta}(a, b) = \frac{1}{2} \parens*{\frac{\delta_{\theta}^{2}(a, b)}{\rho_{\theta}^{2}(b)} + \partial_{a} \delta_{\theta}(a, b)}.
\end{gather}
We now change the dominating measure to $\meas{M}|(x_{\braces{\dot{\tau}, \ddot{\tau}}}, y_{\dot{\tau}}, \theta) \times \leb$, such that the Lebesgue-dominated density $\pi(x_{\ddot{\tau}} | x_{\dot{\tau}}, y_{\dot{\tau}}, \theta)$ becomes the marginal. By the definition of conditional probability, we note that
\begin{gather}
  \der{\meas{X}|(x_{\dot{\tau}}, y_{\dot{\tau}}, \theta)}{\meas{X}|(x_{\braces{\dot{\tau}, \ddot{\tau}}}, y_{\dot{\tau}}, \theta)}(x_{(\dot{\tau}, \ddot{\tau}]}) = \pi(x_{\ddot{\tau}} | x_{\dot{\tau}}, y_{\dot{\tau}}, \theta), \\
  \der{\meas{M}|(x_{\dot{\tau}}, y_{\dot{\tau}}, \theta)}{\meas{M}|(x_{\braces{\dot{\tau}, \ddot{\tau}}}, y_{\dot{\tau}}, \theta)}(x_{(\dot{\tau}, \ddot{\tau}]}) = \op{N}{x_{\ddot{\tau}}; x_{\dot{\tau}}, (\ddot{\tau} - \dot{\tau}) \rho_{\theta}^{2}(y_{\dot{\tau}})},
\end{gather}
and combining both expressions,
\begin{equation}
  \begin{split}
    & \pi(x_{\ddot{\tau}} | x_{\dot{\tau}}, y_{\dot{\tau}}, \theta) \der{\meas{X}|(x_{\braces{\dot{\tau}, \ddot{\tau}}}, y_{\dot{\tau}}, \theta)}{\meas{M}|(x_{\braces{\dot{\tau}, \ddot{\tau}}}, y_{\dot{\tau}}, \theta)}(x_{(\dot{\tau}, \ddot{\tau})}) \\
    & \quad = \op{N}{x_{\ddot{\tau}}; x_{\dot{\tau}}, (\ddot{\tau} - \dot{\tau}) \rho_{\theta}^{2}(y_{\dot{\tau}})} \der{\meas{X}|(x_{\dot{\tau}}, y_{\dot{\tau}}, \theta)}{\meas{M}|(x_{\dot{\tau}}, y_{\dot{\tau}}, \theta)}(x_{(\dot{\tau}, \ddot{\tau}]}).
  \end{split}
\end{equation}
We now define the \emph{complete transition density}
\begin{equation}
  \pi(x_{(\dot{\tau}, \ddot{\tau}]} | x_{\dot{\tau}}, y_{\dot{\tau}}, \theta) = \pi(x_{\ddot{\tau}} | x_{\dot{\tau}}, y_{\dot{\tau}}, \theta) \der{\meas{X}|(x_{\braces{\dot{\tau}, \ddot{\tau}}}, y_{\dot{\tau}}, \theta)}{\meas{M}|(x_{\braces{\dot{\tau}, \ddot{\tau}}}, y_{\dot{\tau}}, \theta)}(x_{(\dot{\tau}, \ddot{\tau})}),
\end{equation}
and changing variables back to $V_{\ddot{\tau}}$, the expression is
\begin{equation}
  \begin{aligned}
    \pi(x_{(\dot{\tau}, \ddot{\tau})}, v_{\ddot{\tau}} | v_{\dot{\tau}}, y_{\dot{\tau}}, \theta)
    & = |\eta_{\theta}'(v_{\ddot{\tau}})| \op{N}{\eta_{\theta}(v_{\ddot{\tau}}); \eta_{\theta}(v_{\dot{\tau}}), (\ddot{\tau} - \dot{\tau}) \rho_{\theta}^{2}(y_{\dot{\tau}})} \\
    & \qquad \times \der{\meas{X}|(X_{\dot{\tau}} = \eta_{\theta}(v_{\dot{\tau}}), y_{\dot{\tau}}, \theta)}{\meas{M}|(X_{\dot{\tau}} = \eta_{\theta}(v_{\dot{\tau}}), y_{\dot{\tau}}, \theta)}(x_{(\dot{\tau}, \ddot{\tau})}, \eta_{\theta}(v_{\ddot{\tau}})).
  \end{aligned}
\end{equation}
We note that for distinct values $\theta \neq \theta^{\dagger}$, $\meas{M}|(X_{\dot{\tau}} = \eta_{\theta}(v_{\dot{\tau}}), y_{\dot{\tau}}, \theta)$ and $\meas{M}|(X_{\dot{\tau}} = \eta_{\theta^{\dagger}}(v_{\dot{\tau}}), y_{\dot{\tau}}, \theta^{\dagger})$ are mutually singular, and therefore $\pi(x_{(\dot{\tau}, \ddot{\tau})}, v_{\ddot{\tau}} | v_{\dot{\tau}}, y_{\dot{\tau}}, \theta)$ and $\pi(x_{(\dot{\tau}, \ddot{\tau})}, v_{\ddot{\tau}} | v_{\dot{\tau}}, y_{\dot{\tau}}, \theta^{\dagger})$ are mutually singular as well. Hence, a noncentered parameterization is required.

The second step is to change variables from the centered bridges $X_{(\dot{\tau}, \ddot{\tau})}$ to noncentered, a priori independent bridges. We define
\begin{equation}
  \zeta_{\theta}(x_{t}; y_{\dot{\tau}}, v_{\braces{\dot{\tau}, \ddot{\tau}}}) = \braces*{x_{t} - \eta_{\theta}(v_{\dot{\tau}}) - (\eta_{\theta}(v_{\ddot{\tau}}) - \eta_{\theta}(v_{\dot{\tau}})) \frac{t - \dot{\tau}}{\ddot{\tau} - \dot{\tau}}} / \rho_{\theta}(y_{\dot{\tau}}), \quad (t \in (\dot{\tau}, \ddot{\tau}))
\end{equation}
and let $\zeta_{\theta}^{-1}$ be the inverse in the first argument:
\begin{equation}
  \zeta_{\theta}^{-1}(z_{t}; y_{\dot{\tau}}, v_{\braces{\dot{\tau}, \ddot{\tau}}}) = \eta_{\theta}(v_{\dot{\tau}}) + (\eta_{\theta}(v_{\ddot{\tau}}) - \eta_{\theta}(v_{\dot{\tau}})) \frac{t - \dot{\tau}}{\ddot{\tau} - \dot{\tau}} + \rho_{\theta}(y_{\dot{\tau}}) z_{t}. \quad (t \in (\dot{\tau}, \ddot{\tau}))
\end{equation}
We change variables to $Z_{(\dot{\tau}, \ddot{\tau})} = \zeta_{\theta}(X_{(\dot{\tau}, \ddot{\tau})}; y_{\dot{\tau}}, v_{\braces{\dot{\tau}, \ddot{\tau}}})$ and note that under $\meas{M}|(x_{\dot{\tau}}, y_{\dot{\tau}}, \theta)$, $Z_{(\dot{\tau}, \ddot{\tau})}$ is a Brownian bridge spanning the origin at times $(\dot{\tau}, \ddot{\tau})$. We further define $\meas{Z}|(x_{\braces{\dot{s}, \ddot{s}}}, y_{\dot{\tau}}, \theta)$ and $\meas{B}_{(\dot{\tau}, \ddot{\tau})}$ as the pushforward measures induced by $Z_{(\dot{\tau}, \ddot{\tau})}$ under $\meas{X}|(x_{\braces{\dot{\tau}, \ddot{\tau}}}, y_{\dot{\tau}}, \theta)$ and $\meas{W}|(x_{\braces{\dot{\tau}, \ddot{\tau}}}, y_{\dot{\tau}}, \theta)$, respectively. Probabilities being conserved under a change of variable, we find that
\begin{equation}
  \begin{split}
    \der{\meas{Z}|(x_{\braces{\dot{\tau}, \ddot{\tau}}}, y_{\dot{\tau}}, \theta)}{\meas{B}_{(\dot{\tau}, \ddot{\tau})}}(z_{(\dot{\tau}, \ddot{\tau})})
    & = \der{\meas{X}|(x_{\braces{\dot{\tau}, \ddot{\tau}}}, y_{\dot{\tau}}, \theta)}{\meas{M}|(x_{\braces{\dot{\tau}, \ddot{\tau}}}, y_{\dot{\tau}}, \theta)} \circ \zeta_{\theta}^{-1}(z_{(\dot{\tau}, \ddot{\tau})}; y_{\dot{\tau}}, v_{\braces{\dot{\tau}, \ddot{\tau}}}) \\
    & = \frac{\op{N}{x_{\ddot{\tau}}; x_{\dot{\tau}}, (\ddot{\tau} - \dot{\tau}) \rho_{\theta}^{2}(y_{\dot{\tau}})}}{\pi(x_{\ddot{\tau}} | x_{\dot{\tau}}, y_{\dot{\tau}}, \theta)} \\
    & \qquad \times \der{\meas{X}|(x_{\dot{\tau}}, y_{\dot{\tau}}, \theta)}{\meas{W}|(x_{\dot{\tau}}, y_{\dot{\tau}}, \theta)}(\zeta_{\theta}^{-1}(z_{(\dot{\tau}, \ddot{\tau})}; y_{\dot{\tau}}, v_{\braces{\dot{\tau}, \ddot{\tau}}}), x_{\ddot{\tau}}),
  \end{split}
\end{equation}
which, in conjunction with $\pi(x_{(\dot{\tau}, \ddot{\tau})}, v_{\ddot{\tau}} | v_{\dot{\tau}}, y_{\dot{\tau}}, \theta)$, gives us the noncentered complete transition density:
\begin{equation}
  \begin{split}
    \pi(z_{(\dot{\tau},\ddot{\tau})}, v_{\ddot{\tau}} | v_{\dot{\tau}}, y_{\dot{\tau}}, \theta)
    & = |\eta_{\theta}'(v_{\ddot{\tau}})| \op{N}{\eta_{\theta}(v_{\ddot{\tau}}); \eta_{\theta}(v_{\dot{\tau}}), (\ddot{\tau} - \dot{\tau}) \rho_{\theta}^{2}(y_{\dot{\tau}})} \\
    & \qquad \times \der{\meas{X}|(X_{\dot{\tau}} = \eta_{\theta}(v_{\dot{\tau}}), y_{\dot{\tau}}, \theta)}{\meas{M}|(X_{\dot{\tau}} = \eta_{\theta}(v_{\dot{\tau}}), y_{\dot{\tau}}, \theta)}(\zeta_{\theta}^{-1}(z_{(\dot{\tau},\ddot{\tau})}; y_{\dot{\tau}}, v_{\braces{\dot{\tau}, \ddot{\tau}}}), \eta_{\theta}(v_{\ddot{\tau}})) \\
    & = |\eta_{\theta}'(v_{\ddot{\tau}})| \op{N}{\eta_{\theta}(v_{\ddot{\tau}}); \eta_{\theta}(v_{\dot{\tau}}), (\ddot{\tau} - \dot{\tau})  \rho_{\theta}^{2}(y_{\dot{\tau}})} \\
    & \qquad \times \exp\braces*{\rho_{\theta}^{-2}(y_{\dot{\tau}}) (\Delta_{\theta}(\eta_{\theta}(v_{\ddot{\tau}}), y_{\dot{\tau}}) - \Delta_{\theta}(\eta_{\theta}(v_{\dot{\tau}}), y_{\dot{\tau}}))} \\
    & \qquad \times \exp\braces*{-\int_{\dot{\tau}}^{\ddot{\tau}} \varphi_{\theta}(\zeta_{\theta}^{-1}(z_{t}; y_{\dot{\tau}}, v_{\braces{\dot{\tau}, \ddot{\tau}}}), y_{\dot{\tau}}) \dd{t}},
  \end{split}
\end{equation}
where the dominating measure is $\meas{B}_{(\dot{\tau}, \ddot{\tau})} \times \leb$, and by construction,
\begin{equation}
  \int \pi(z_{(\dot{\tau},\ddot{\tau})}, v_{\ddot{\tau}} | v_{\dot{\tau}}, y_{\dot{\tau}}, \theta) \meas{B}_{(\dot{\tau}, \ddot{\tau})}(\dd{z_{(\dot{\tau},\ddot{\tau})}}) = \pi(v_{\ddot{\tau}} | v_{\dot{\tau}}, y_{\dot{\tau}}, \theta).
\end{equation}

\section{Localizing and Adapting the Hidden Data Update}
\label{sec:localize}

The simple independence proposal described in the main text will tend to break down in various ways as data accrues. We consider the \emph{infill} regime, where $\op{mesh} s \to 0$, and the \emph{outfill} regime, where $|s|, \omega \to \infty$, and develop strategies to maintain good computational performance in both. Both in the infill and the outfill regime, the independence proposal from $\pi(y^{\dagger} | \lambda)$ will be an increasingly bad fit for the full conditional, causing a degradation in the acceptance probability and slow mixing of the algorithm. Similarly, the independence proposal from $\meas{M}|(x_{s}, y^{\dagger}, \theta)$ becomes a bad fit as the time interval between observation increases, or in the presence of transitions in $y$ and the associated discontinuities in the drift function. To those obstacles we add the aforementioned phenomenon of the exponential slowdown of the 2-coin algorithm as the time horizon recedes. Thus, we devise a localized, scalable $(V_{R}, Z, Y)$-update that addresses both the infill and the outfill asymptotic regime.

The approach consists of conditioning the $(V_{R}, Z, Y)$-update on $Y_{N}$, where $N \subseteq (s \setdiff \braces{0, \omega})$ is a random set of times. If no element of $s$ is included in $N$ with probability 1, this update may be thought of as a \emph{random scan} Gibbs update, which preserves ergodicity of the Markov chain. The main rationale to limiting conditioning times to elements of $s$ is that in its EA2/EA3 representation, $Z$ is only semi-Markovian at times $\tau$, and therefore the full conditional does not factorize neatly at times $\nu \not\subset \tau$.

With $\nu$ fixed, we generate updates according to $\pi(v_{r}, y \setdiff y_{\nu} | v_{s}, y_{\nu}, \theta, \lambda)$. On the one hand, the conditional proposal $\kappa(v_{r}, y \setdiff y_{\nu} | v_{s}, y_{\nu})$ has a smaller step size, thereby increasing the acceptance probability. On the other hand, by the Markov property, we immediately benefit from the factorization
\begin{gather}
  \begin{split}
    \kappa(v_{r^{\dagger}}^{\dagger}, z^{\dagger}, y^{\dagger} | v_{s}, y_{\nu})
    & = \prod_{(\dot{\nu} \sim \ddot{\nu}) \in \nu} \kappa(v_{r^{\dagger} \cap (\dot{\nu}, \ddot{\nu})}^{\dagger}, z_{(\dot{\nu}, \ddot{\nu})}^{\dagger}, y_{(\dot{\nu}, \ddot{\nu})}^{\dagger} | v_{s}, y_{\braces{\dot{\nu}, \ddot{\nu}}}) \\
    & \quad \times \kappa(v_{r^{\dagger} \cap (0, \nu_{1})}^{\dagger}, z_{(0, \nu_{1})}^{\dagger}, y_{[0, \nu_{1})}^{\dagger} | v_{s}, y_{\nu_{1}}) \\
    & \quad \times \kappa(v_{r^{\dagger} \cap (\nu_{|\nu|}, \omega)}^{\dagger}, z_{(\nu_{|\nu|}, \omega)}^{\dagger}, y_{(\nu_{|\nu|}, \omega]}^{\dagger} | v_{s}, y_{\nu_{|\nu|}}),
  \end{split}
\end{gather}
with an analogous factorization for $\pi(v_{r}, z, y \setdiff y_{\nu} | v_{s}, y_{\nu}, \theta, \lambda)$, so generation and acceptance of the proposal may be partitioned according to $\nu$. This further increases the acceptance probability, and reduces Bernoulli factory run time. The proposal law
\begin{equation}
  \begin{split}
    & \kappa(v_{r^{\dagger} \cap (\dot{\nu}, \ddot{\nu})}^{\dagger}, z_{(\dot{\nu}, \ddot{\nu})}^{\dagger}, y_{(\dot{\nu}, \ddot{\nu})}^{\dagger} | v_{s}, y_{\braces{\dot{\nu}, \ddot{\nu}}}) \\
    & \quad = \pi(y_{(\dot{\nu}, \ddot{\nu})}^{\dagger} | y_{\braces{\dot{\nu}, \ddot{\nu}}}, \lambda) \kappa(v_{r^{\dagger} \cap (\dot{\nu}, \ddot{\nu})}^{\dagger} | v_{s}, y^{\dagger}) \prod_{(\dot{\tau} \sim \ddot{\tau}) \in \tau^{\dagger} \cap [\dot{\nu}, \ddot{\nu}]} \kappa(z_{(\dot{\tau}, \ddot{\tau})}^{\dagger})
  \end{split}
\end{equation}
involves the Markov jump process bridge law $\pi(y_{(\dot{\nu}, \ddot{\nu})}^{\dagger} | y_{\braces{\dot{\nu}, \ddot{\nu}}}, \lambda)$, while the edge proposals
\begin{gather}
  \begin{split}
    & \kappa(v_{r^{\dagger} \cap (0, \nu_{1})}^{\dagger}, z_{(0, \nu_{1})}^{\dagger}, y_{[0, \nu_{1})}^{\dagger} | v_{s}, y_{\nu_{1}}) \\
    & \quad = \pi(y_{[0, \nu_{1})}^{\dagger} | y_{\nu_{1}}, \lambda) \kappa(v_{r^{\dagger} \cap (0, \nu_{1})}^{\dagger} | v_{s}, y^{\dagger}) \prod_{(\dot{\tau} \sim \ddot{\tau}) \in \tau^{\dagger} \cap [0, \nu_{1}]} \kappa(z_{(\dot{\tau}, \ddot{\tau})}^{\dagger}), \\
  \end{split} \\
  \begin{split}
    & \kappa(v_{r^{\dagger} \cap (\nu_{|\nu|}, \omega)}^{\dagger}, z_{(\nu_{|\nu|}, \omega)}^{\dagger}, y_{(\nu_{|\nu|}, \omega]}^{\dagger} | v_{s}, y_{\nu_{|\nu|}}) \\
    & \quad = \pi(y_{(\nu_{|\nu|}, \omega]}^{\dagger} | y_{\nu_{|\nu|}}, \lambda) \kappa(v_{r^{\dagger} \cap (\nu_{|\nu|}, \omega)}^{\dagger} | v_{s}, y^{\dagger}) \prod_{(\dot{\tau} \sim \ddot{\tau}) \in \tau^{\dagger} \cap [\nu_{|\nu|}, \omega]} \kappa(z_{(\dot{\tau}, \ddot{\tau})}^{\dagger}),
  \end{split}
\end{gather}
merely involve the backward and forward law $\pi(y_{[0, \nu_{1})}^{\dagger} | y_{\nu_{1}}, \lambda)$ and $\pi(y_{(\nu_{|\nu|}, \omega]}^{\dagger} | y_{\nu_{|\nu|}}, \lambda)$ respectively. Simulation of the Markov jump process bridges from $\pi(y_{(\dot{\nu}, \ddot{\nu})}^{\dagger} | y_{\braces{\dot{\nu}, \ddot{\nu}}}, \lambda)$ may be carried out according to any of the schemes proposed in \citep{hobolth2009simulation}. We further observe the decomposition
\begin{equation}
  \kappa(v_{r^{\dagger} \cap (\dot{\nu}, \ddot{\nu})}^{\dagger} | v_{s}, y^{\dagger}) = \prod_{(\dot{s} \sim \ddot{s}) \in s \cap [\dot{\nu}, \ddot{\nu}]} \kappa(v_{r^{\dagger} \cap (\dot{s}, \ddot{s})}^{\dagger} | v_{\braces{\dot{s}, \ddot{s}}}, y_{[\dot{s}, \ddot{s}]}^{\dagger}), \qquad ((\dot{\nu} \sim \ddot{\nu}) \in \nu \cup \braces{0, \omega})
\end{equation}
where $\kappa(v_{r^{\dagger} \cap (\dot{s}, \ddot{s})}^{\dagger} | v_{\braces{\dot{s}, \ddot{s}}}, y_{[\dot{s}, \ddot{s}]}^{\dagger})$ is given by \eqref{eq:hidpropdens}, and samples are obtained as in Algorithm \ref{alg:indprop}. Having constructed the proposal, we now observe that the acceptance odds factorize spontaneously at times $\varsigma = (s \cap \braces{t: y_{t} = y_{t}^{\dagger}}) \supseteq \nu$. Noting that since $y_{\varsigma} = y_{\varsigma}^{\dagger}$, the acceptance probability satisfies
\begin{equation}
  \begin{split}
    \frac{\alpha_{(V_{R}, Z, Y | y_{\nu})}(\braces{v_{r^{\dagger}}^{\dagger}, z^{\dagger}, y^{\dagger}}, \braces{v_{r}, z, y})}{\alpha_{(V_{R}, Z, Y | y_{\nu})}(\braces{v_{r}, z, y}, \braces{v_{r^{\dagger}}^{\dagger}, z^{\dagger}, y^{\dagger}})}
    & = \frac{\kappa(v_{r}, z, y | v_{s}, y_{\nu})}{\kappa(v_{r^{\dagger}}^{\dagger}, z^{\dagger}, y^{\dagger} | v_{s}, y_{\nu})} \frac{\pi(v_{r^{\dagger}}^{\dagger}, z^{\dagger}, y^{\dagger} | v_{s}, y_{\nu}, \theta, \lambda)}{\pi(v_{r}, z, y | v_{s}, y_{\nu}, \theta, \lambda)} \\
    & = \frac{\kappa(v_{r}, z, y | v_{s}, y_{\varsigma})}{\kappa(v_{r^{\dagger}}^{\dagger}, z^{\dagger}, y^{\dagger} | v_{s}, y_{\varsigma})} \frac{\pi(v_{r^{\dagger}}^{\dagger}, z^{\dagger}, y^{\dagger} | v_{s}, y_{\varsigma}, \theta, \lambda)}{\pi(v_{r}, z, y | v_{s}, y_{\varsigma}, \theta, \lambda)} \\
    & = \frac{\alpha_{(V_{R}, Z, Y | y_{\varsigma})}(\braces{v_{r^{\dagger}}^{\dagger}, z^{\dagger}, y^{\dagger}}, \braces{v_{r}, z, y})}{\alpha_{(V_{R}, Z, Y | y_{\varsigma})}(\braces{v_{r}, z, y}, \braces{v_{r^{\dagger}}^{\dagger}, z^{\dagger}, y^{\dagger}})}.
  \end{split}
\end{equation}
Then, defining $\gamma_{(\dot{\varsigma}, \ddot{\varsigma})}^{\dagger} = \braces{v_{r^{\dagger} \cap (\dot{\varsigma}, \ddot{\varsigma})}^{\dagger}, z_{(\dot{\varsigma}, \ddot{\varsigma})}^{\dagger}, y_{[\dot{\varsigma}, \ddot{\varsigma}]}^{\dagger}}$ and applying the Markov property, 
\begin{equation}
  \begin{split}
    & \frac{\alpha_{(V_{R}, Z, Y | y_{\varsigma})}(\braces{v_{r^{\dagger}}^{\dagger}, z^{\dagger}, y^{\dagger}}, \braces{v_{r}, z, y})}{\alpha_{(V_{R}, Z, Y | y_{\varsigma})}(\braces{v_{r}, z, y}, \braces{v_{r^{\dagger}}^{\dagger}, z^{\dagger}, y^{\dagger}})} \\
    & \quad = \prod_{(\dot{\varsigma} \sim \ddot{\varsigma}) \in \varsigma \cap \braces{0, \omega}} \frac{\kappa(\gamma_{(\dot{\varsigma}, \ddot{\varsigma})} | v_{s}, y_{\varsigma})}{\kappa(\gamma_{(\dot{\varsigma}, \ddot{\varsigma})}^{\dagger} | v_{s}, y_{\varsigma})} \frac{\pi(\gamma_{(\dot{\varsigma}, \ddot{\varsigma})}^{\dagger} | v_{s}, y_{\varsigma}, \theta, \lambda)}{\pi(\gamma_{(\dot{\varsigma}, \ddot{\varsigma})}^{\dagger} | v_{s}, y_{\varsigma}, \theta, \lambda)} \\
    & \quad = \prod_{(\dot{\varsigma} \sim \ddot{\varsigma}) \in \varsigma \cap \braces{0, \omega}} \frac{\alpha_{(V_{R}, Z, Y | y_{\varsigma})}(\gamma_{(\dot{\varsigma}, \ddot{\varsigma})}^{\dagger}, \gamma_{(\dot{\varsigma}, \ddot{\varsigma})})}{\alpha_{(V_{R}, Z, Y | y_{\varsigma})}(\gamma_{(\dot{\varsigma}, \ddot{\varsigma})}, \gamma_{(\dot{\varsigma}, \ddot{\varsigma})}^{\dagger})},
  \end{split}
\end{equation}
so each section is accepted or rejected independently with odds
\begin{equation}
  \prod_{(\dot{s} \sim \ddot{s}) \in s \cap [\dot{\varsigma}, \ddot{\varsigma}]} \frac{\kappa(v_{r \cap (\dot{s}, \ddot{s})} | v_{\braces{\dot{s}, \ddot{s}}}, y_{[\dot{s}, \ddot{s}]})}{\kappa(v_{r^{\dagger} \cap (\dot{s}, \ddot{s})}^{\dagger} | v_{\braces{\dot{s}, \ddot{s}}}, y_{[\dot{s}, \ddot{s}]}^{\dagger})} \frac{\prod_{(\dot{\tau} \sim \ddot{\tau}) \in \tau^{\dagger} \cap [\dot{\varsigma}, \ddot{\varsigma}]} \pi(z_{(\dot{\tau}, \ddot{\tau})}^{\dagger}, v_{\ddot{\tau}}^{\dagger} | v_{\dot{\tau}}^{\dagger}, y_{\dot{\tau}}^{\dagger}, \theta)}{\prod_{(\dot{\tau} \sim \ddot{\tau}) \in \tau \cap [\dot{\varsigma}, \ddot{\varsigma}]} \pi(z_{(\dot{\tau}, \ddot{\tau})}, v_{\ddot{\tau}} | v_{\dot{\tau}}, y_{\dot{\tau}}, \theta)}.
\end{equation}
As for adaptation, $\nu$ may be picked in various ways, as long as each element of $s$ is included with probability less than 1. We propose including elements of $s$ independently, with inclusion probability chosen such that the local acceptance rate reaches a target rate. For $\dot{s} \in s$, We define the local acceptance probability as
\begin{equation}
  \begin{dcases}
    \alpha_{(V_{R}, Z, Y | y_{\varsigma})}(\gamma_{(\dot{\varsigma}, \ddot{\varsigma})}^{\dagger}, \gamma_{(\dot{\varsigma}, \ddot{\varsigma})}) & (\text{if $\dot{s} \in (\dot{\varsigma}, \ddot{\varsigma})$}) \\
    (\alpha_{(V_{R}, Z, Y | y_{\varsigma})}(\gamma_{(\dot{\varsigma}, \ddot{\varsigma})}^{\dagger}, \gamma_{(\dot{\varsigma}, \ddot{\varsigma})}) + \alpha_{(V_{R}, Z, Y | y_{\varsigma})}(\gamma_{(\ddot{\varsigma}, \dddot{\varsigma})}^{\dagger}, \gamma_{(\ddot{\varsigma}, \dddot{\varsigma})})) / 2 & (\text{if $\dot{s} = \ddot{\varsigma}$})
  \end{dcases}
\end{equation}
where in the latter instance, $(\dot{\varsigma}, \ddot{\varsigma})$ and $(\ddot{\varsigma}, \dddot{\varsigma})$ are the pairs in $\varsigma$ neighboring $\ddot{\varsigma}$. An adaptive MCMC method such as Adapting increasingly rarely may then be used to adjust $\op{Pr}{\dot{s} \in N}$ such that the local acceptance probability hits its target value on average.



\section{Approximate MCMC Algorithm}
\label{sec:approx}

When pre-adapting the main sampling or optimization run with an approximate model, we use the first-order Euler-Maruyama scheme, which replaces the intractable transition density with the linearized approximation $\est{\pi}(v_{\ddot{\tau}} | v_{\dot{\tau}}, y_{\dot{\tau}}, \theta)$. We can then run a Gibbs sampler with $(V_{R}, Y)$- and $(\Theta, \Lambda)$-updates, or an MCEM algorithm with E-Step over $(V_{R}, Y)$ and M-step over $(\Theta, \Lambda)$. It is usually preferable to linearize the reduced diffusion $X$, resulting in the transition density approximation
\begin{equation}
  \est{\pi}(v_{\ddot{\tau}} | v_{\dot{\tau}}, y_{\dot{\tau}}, \theta) = |\eta_{\theta}'(v_{\ddot{\tau}})| \op{N}{x_{\ddot{\tau}}; x_{\dot{\tau}} + (\ddot{\tau} - \dot{\tau}) \delta_{\theta}(x_{\dot{\tau}}), (\ddot{\tau} - \dot{\tau}) \rho_{\theta}^{2}(y_{\dot{\tau}})},
\end{equation}
which coincides with the higher-order Milstein approximation. The approximation bias can be reduced by imputing additional observations. If we impute observations at times $u_{(\dot{\tau}, \ddot{\tau})} \subset (\dot{\tau}, \ddot{\tau})$ and define $\est{x}_{(\dot{\tau}, \ddot{\tau})} = x_{u_{(\dot{\tau}, \ddot{\tau})}}$, the refined approximate likelihood is
\begin{gather}
  \est{\pi}(v_{\ddot{\tau}}, \est{x}_{(\dot{\tau}, \ddot{\tau})} | v_{\dot{\tau}}, y_{\dot{\tau}}, \theta) = \bars{\eta_{\theta}'(v_{\ddot{\tau}})} \prod_{(\dot{u} \sim \ddot{u}) \in u_{(\dot{\tau}, \ddot{\tau})}} \op{N}{x_{\ddot{\tau}}; \substack{x_{\dot{\tau}} + (\ddot{\tau} - \dot{\tau}) \delta_{\theta}(x_{\dot{\tau}}), \\ (\ddot{\tau} - \dot{\tau}) \rho_{\theta}^{2}(y_{\dot{\tau}})}}, \\
  \op*{lim}[\op{mesh} u_{(\dot{\tau}, \ddot{\tau})} \to 0] \op{E}[\est{X}_{(\dot{\tau}, \ddot{\tau})}]{\est{\pi}(v_{\ddot{\tau}}, \est{X}_{(\dot{\tau}, \ddot{\tau})} | v_{\dot{\tau}}, y_{\dot{\tau}}, \theta) | v_{\braces{\dot{\tau}, \ddot{\tau}}}, y_{\dot{\tau}}, \theta} = \pi(v_{\ddot{\tau}} | v_{\dot{\tau}}, y_{\dot{\tau}}, \theta).
\end{gather}
where the latter is due do weak convergence of the Milstein approximation. Conversely, at higher imputation rates, such an imputation scheme negatively affects mixing, as examined in detail in \citep{roberts2001inference}. As in the exact algorithm, we address this by switching to the non-centered parameterization. Defining $\est{z}_{(\dot{\tau}, \ddot{\tau})} = \zeta_{\theta}(\est{x}_{(\dot{\tau}, \ddot{\tau})}; y_{\dot{\tau}}, v_{\braces{\dot{\tau}, \ddot{\tau}}})$,
\begin{equation}
\begin{split}
  & \est{\pi}(v_{\ddot{\tau}}, \est{z}_{(\dot{\tau}, \ddot{\tau})} | v_{\dot{\tau}}, y_{\dot{\tau}}, \theta) \\
  & \quad = \bars{\eta_{\theta}'(v_{\ddot{\tau}})} \prod_{z_{t} \in \est{z}_{(\dot{\tau}, \ddot{\tau})}} \bars{\partial_{z_{t}} \zeta_{\theta}^{-1}(z_{t}; y_{\dot{\tau}}, v_{\braces{\dot{\tau}, \ddot{\tau}}})} \\
  & \qquad \times \prod_{(\dot{u} \sim \ddot{u}) \in u_{(\dot{\tau}, \ddot{\tau})}} \op{N}{\substack{\zeta_{\theta}^{-1}(z_{\ddot{u}}; y_{\dot{\tau}}, v_{\braces{\dot{\tau}, \ddot{\tau}}}); \\ \zeta_{\theta}^{-1}(z_{\dot{u}}; y_{\dot{\tau}}, v_{\braces{\dot{\tau}, \ddot{\tau}}}) + (\ddot{u} - \dot{u}) \delta_{\theta}(\zeta_{\theta}^{-1}(z_{\dot{u}}; y_{\dot{\tau}}, v_{\braces{\dot{\tau}, \ddot{\tau}}}), y_{\dot{\tau}}, v_{\braces{\dot{\tau}, \ddot{\tau}}}), \\ (\ddot{u} - \dot{u}) \rho_{\theta}^{2}(y_{\dot{\tau}})}},
\end{split}
\end{equation}
where we slightly abuse notation by setting $\zeta_{\theta}^{-1}(z_{\dot{\tau}}; y_{\dot{\tau}}, v_{\braces{\dot{\tau}, \ddot{\tau}}}) = x_{\dot{\tau}}$ and $\zeta_{\theta}^{-1}(z_{\ddot{\tau}}; y_{\dot{\tau}}, v_{\braces{\dot{\tau}, \ddot{\tau}}}) = x_{\ddot{\tau}}$, and the Jacobian is given by
\begin{equation}
  \bars{\partial_{z_{t}} \zeta_{\theta}^{-1}(z_{t}; y_{\dot{\tau}}, v_{\braces{\dot{\tau}, \ddot{\tau}}})} = \rho_{\theta}(y_{\dot{\tau}}). \qquad (t \in (\dot{\tau}, \ddot{\tau}))
\end{equation}
This parameterization of the missing data conserves ergodicity as $\op{mesh} u_{(\dot{\tau}, \ddot{\tau})} \to 0$, and gives us a viable, approximate augmentation scheme within the same Gibbs blocking scheme as in the exact algorithm. The approximate posterior targeted by that sampler is
\begin{equation}
  \est{\pi}(v_{r}, \est{z}, y, \theta, \lambda | v_{s}) \propto \pi(\theta) \pi(\lambda) \prod_{(\dot{\tau} \sim \ddot{\tau}) \in \tau} \est{\pi}(v_{\ddot{\tau}}, \est{z}_{(\dot{\tau}, \ddot{\tau})} | v_{\dot{\tau}}, y_{\dot{\tau}}, \theta) \pi(y_{\ddot{\tau}} | y_{\dot{\tau}}, \lambda),
\end{equation}
and its Gibbs updates are
\begin{align}
  (V_{R}, \est{Z}, Y): \quad & \est{\pi}(v_{r}, \est{z}, y | v_{s}, \theta, \lambda) \propto \prod_{(\dot{\tau} \sim \ddot{\tau}) \in \tau} \est{\pi}(\est{z}_{(\dot{\tau}, \ddot{\tau})}, v_{\ddot{\tau}} | v_{\dot{\tau}}, y_{\dot{\tau}}, \theta) \pi(y_{\ddot{\tau}} | y_{\dot{\tau}}, \lambda), \\
  \Theta: \quad & \est{\pi}(\theta | v_{\tau}, \est{z}, y) \propto \pi(\theta) \prod_{(\dot{\tau} \sim \ddot{\tau}) \in \tau} \est{\pi}(\est{z}_{(\dot{\tau}, \ddot{\tau})}, v_{\ddot{\tau}} | v_{\dot{\tau}}, y_{\dot{\tau}}, \theta), \\
  \lambda: \quad & \pi(\lambda | y) \propto \pi(\lambda) \pi(y | \lambda),
\end{align}
where we propose $\est{z}_{(\dot{\tau}, \ddot{\tau})}$ according to $\meas{B}_{(\dot{\tau}, \ddot{\tau})}$. Note that in this instance, Barker-within-Gibbs updates can be replaced by conventional Metropolis-within-Gibbs updates. Nonetheless, for the purpose of warm-starting an exact algorithm, it may be preferable to use Barker-within-Gibbs updates, as this results in a smaller, more appropriate step size at the start of the exact run.

\section{Poisson Coin Algorithm}
\label{sec:poisson}

An essential ingredient to exact diffusion inference is the ability to simulate coins of probability
\begin{equation}
  \exp\braces*{\int_{0}^{\omega} (\lb{f} - f_{t}) \dd{t}}
\end{equation}
for various paths $f: [0, \omega] \to [\lb{f}, \ub{f}]$. This is addressed by the \emph{Poisson coin} algorithm of \citep{beskos2005exact}. Notice that $f$ has to be upper bounded at $\ub{f}$ in order to implement the Poisson coin algorithm. The main insight is that if we can construct and assess a tractable event $E$ such that $\op{Pr}{E} = p$, evaluation of $p$ is not necessary to flipping $p$-coins. We recall that a \emph{homogeneous Poisson process} $\Psi$ on $\reals^{d}$ is defined as a point process which satisfies
\begin{equation}
  |\Psi \cap B| \sim \op{Pois}{\lambda \times \op{Vol}{B}}
\end{equation}
for every bounded set $B \in \reals^{d}$, where $\lambda$ is the rate of the process. Moreover, $\Psi \cap B$ is again a Poisson process. We use the shorthand $\op{PP}{B, \lambda}$ for a rate $\lambda$ Poisson process on $B$. We further define the epigraph of $t \mapsto f_{t} - \lb{f}$ as
\begin{equation}
  \op{epi}{f - \lb{f}} = \braces{(t, a) \in [0, \omega] \times [0, \infty): a \leq f_{t} - \lb{f}},
\end{equation}
and notice that is has area $\int_{0}^{\omega} (\lb{f} - f_{t}) \dd{t}$. Furthermore, let $\Psi \sim \op{PP}{[0, \omega] \times [0, \ub{f} - \lb{f}], 1}$, and notice that since $\op{epi}{f - \lb{f}} \subset [0, \omega] \times [0, \ub{f} - \lb{f}]$, the intersection $\op{epi}{f - \lb{f}} \cap \Psi$ is a unit rate Poisson process on the epigraph. By definition of the Poisson process,
\begin{gather}
  |\op{epi}{f - \lb{f}} \cap \Psi| \sim \op{Pois}{\int_{0}^{\omega} (f_{t} - \lb{f}) \dd{t}}, \\
  \op{Pr}{|\op{epi}{f - \lb{f}} \cap \Psi| = 0} = \exp\braces*{\int_{0}^{\omega} (\lb{f} - f_{t}) \dd{t}},
\end{gather}
where the latter is a property of the Poisson distribution, and hence $\braces{|(\op{epi} f_{t}) \cap \Psi| = 0}$ is an appropriate choice of $E$. We can assess the event by observing that
\begin{equation}
  \braces{|\op{epi}{f - \lb{f}} \cap \Psi| = 0} = \bigcap_{(T, A) \in \Psi} \braces{A > f_{T} - \lb{f}},
\end{equation}
and since $|\Phi| < \infty$ almost surely, ascertaining the value of the event merely requires evaluating $f$ at a finite number of locations.

We note that the complementary event $\braces{|\op{epi}{f - \lb{f}} \cap \Psi| > 0}$ can often be ascertained without simulating the entire Poisson process, since only one point must fall into the epigraph - indeed, we recommend simulating $\Psi$ in slices from the bottom up, e.g. $[0, \omega] \times [0, (\ub{f} - \lb{f}) / 2]$ and $[0, \omega] \times [(\ub{f} - \lb{f}) / 2, \ub{f} - \lb{f}]$, and checking for each slice whether one of the points falls into the epigraph. If so, the simulation of the remaining slices can be skipped, since the Poisson coin is already known to be 0.

\section{Pseudocode Specification}
\label{sec:pseudocode}

Below we provide a pseudocode specification that consolidates the description of the Gibbs sampler in Section \ref{sec:gibbs} and its generalization in Appendix \ref{sec:localize}. Since their exact implementation goes beyond the scope of this paper, we do not fully specify the primitives $\texttt{construct-hidden-coin}(\theta, y_{\dot{\tau}}, v_{\braces{\dot{\tau}, \ddot{\tau}}})$ and $\texttt{construct-param-coin}(\theta^{\dagger}, \theta, y_{\dot{\tau}}, v_{\braces{\dot{\tau}, \ddot{\tau}}})$. The former returns a function that takes a representation of $z_{(\dot{\tau}, \ddot{\tau})}$ and returns a coin toss of probability $q_{\theta}(z_{(\dot{\tau}, \ddot{\tau})}, y_{\dot{\tau}}, v_{\braces{\dot{\tau}, \ddot{\tau}}})$ and the updated representation of $z_{(\dot{\tau}, \ddot{\tau})}$, including any further information revealed retrospectively. The latter takes a representation of $z_{(\dot{\tau}, \ddot{\tau})}$ and returns coin tosses of probability $\exp\braces{-\int_{\dot{\tau}}^{\ddot{\tau}} \xi_{\theta^{\dagger}, \theta}(z_{t}, y_{\dot{\tau}}, v_{\braces{\dot{\tau}, \ddot{\tau}}}) \dd{t}}$ and the updated representation of $z_{(\dot{\tau}, \ddot{\tau})}$. Their implementation with the Poisson coin algorithm is sketched out in Section \ref{ssec:retro} and Supplement \ref{sec:poisson}, and described in further detail e.g. in \citep{gonccalves2023exact}. The notation below uses the product operator to aggregate multiple coin toss functions into a single coin toss function with probability equal to the product of the constituting probabilities, e.g. $\texttt{construct-hidden-coin}(\theta, y_{\tau_{0}}, v_{\braces{\tau_{0}, \tau_{1}}}) \times \texttt{construct-hidden-coin}(\theta, y_{\tau_{1}}, v_{\braces{\tau_{1}, \tau_{2}}})$ results in a function that takes a representation of $z_{(\tau_{0}, \tau_{1})} \cup z_{(\tau_{1}, \tau_{2})}$ and returns a coin toss of probability $q_{\theta}(z_{(\tau_{0}, \tau_{1})}, y_{\tau_{0}}, v_{\braces{\tau_{0}, \tau_{1}}}) \times q_{\theta}(z_{(\tau_{1}, \tau_{2})}, y_{\tau_{1}}, v_{\braces{\tau_{1}, \tau_{2}}})$ as well as the updated representation.

\begin{breakablealgorithm}
  \caption{Update for $\pi(\theta, \lambda, v_{r}, z, y | v_{s})$ (See Section \ref{sec:gibbs}). \label{alg:gibbs}}
  \begin{algorithmic}
    \Function{update-all}{$\theta, \lambda, v_{\tau}, y, z$}
    \State $v_{r^{*}}^{*}, z^{*}, y^{*} \gets \Call{update-hidden}{\theta, \lambda, v_{r}, v_{s}, y, z}$
    \State $\theta^{*}, z^{*} \gets \Call{update-param}{\theta, v_{\tau^{*}}^{*}, z^{*}, y^{*}}$
    \State $\lambda^{*} \gets \Call{update-generator}{y^{*}}$ \Comment{See Section \ref{ssec:gen}}
    \State \Return $\theta^{*}, \lambda^{*}, v_{r^{*}}^{*}, y^{*}, z^{*}$
    \EndFunction
  \end{algorithmic}
\end{breakablealgorithm}

\begin{breakablealgorithm}
  \caption{Localized Barker-within-Gibbs update for $\pi(v_{r}, z, y | v_{s}, y_{\nu}, \theta, \lambda)$ (see Appendix \ref{sec:localize}). \label{alg:loc}}
  \begin{algorithmic}
    \Function{update-hidden}{$\theta, \lambda, v_{s}, y, y_{\nu}, z$}
    \State $\nu \sim \kappa(\nu)$ \Comment{S.t. $\nu \subseteq s \setdiff \braces{0, \omega}$ and $\op{Pr}{\dot{s} \in \nu} < 1$ for $\dot{s} \in {s}$. For independence sampler, set $\nu = \emptyset$}
    \State $y^{\dagger} \sim \pi(y \setdiff y_{\nu} | y_{\nu}, \lambda)$ \Comment{See \cite{hobolth2009simulation}}
    \For{$(\dot{s} \sim \ddot{s}) \in s$}
      \State $v_{r^{\dagger} \cap (\dot{s}, \ddot{s})}^{\dagger} \sim \kappa(v_{r^{\dagger} \cap (\dot{s}, \ddot{s})}^{\dagger} | v_{\braces{\dot{s}, \ddot{s}}}, y_{[\dot{s}, \ddot{s}]}^{\dagger})$ \Comment{See Algorithm \ref{alg:indprop}}
    \EndFor
    \For{$(\dot{\tau} \sim \ddot{\tau}) \in \tau^{\dagger}$}
      \State $z_{(\dot{\tau}, \ddot{\tau})}^{\dagger} \sim \meas{B}_{(\dot{\tau}, \ddot{\tau})}$
      \State $c_{1}^{(\dot{\tau}, \ddot{\tau})} \gets d_{\theta}(z_{(\dot{\tau}, \ddot{\tau})}^{\dagger}, y_{\dot{\tau}}^{\dagger}, v_{\braces{\dot{\tau}, \ddot{\tau}}}^{\dagger})$
      \State $\varpi_{1}^{(\dot{\tau}, \ddot{\tau})} \gets \Call{construct-hidden-coin}{\theta, y_{\dot{\tau}}^{\dagger}, v_{\braces{\dot{\tau}, \ddot{\tau}}}^{\dagger}}$ \Comment{See Section \ref{ssec:retro}}
    \EndFor
    \For{$(\dot{\tau} \sim \ddot{\tau}) \in \tau$}
      \State $c_{2}^{(\dot{\tau}, \ddot{\tau})} \gets d_{\theta}(z_{(\dot{\tau}, \ddot{\tau})}, y_{\dot{\tau}}, v_{\braces{\dot{\tau}, \ddot{\tau}}})$
      \State $\varpi_{2}^{(\dot{\tau}, \ddot{\tau})} \gets \Call{construct-hidden-coin}{\theta, y_{\dot{\tau}}, v_{\braces{\dot{\tau}, \ddot{\tau}}}}$ \Comment{See Section \ref{ssec:retro}}
    \EndFor
    \For{$(\dot{\nu} \sim \ddot{\nu}) \in s \cap (\braces{0, \omega} \cup \braces{t : y_{t} = y_{t}^{\dagger}})$}
      \State $c_{1} \gets \prod_{(\dot{s} \sim \ddot{s}) \in s \cap [\dot{\nu}, \ddot{\nu}]} \kappa(v_{r \cap (\dot{s}, \ddot{s})} | v_{\braces{\dot{s}, \ddot{s}}}, y_{[\dot{s}, \ddot{s}]}) \prod_{(\dot{\tau} \sim \ddot{\tau}) \in \tau^{\dagger} \cap [\dot{\nu}, \ddot{\nu}]} c_{1}^{(\dot{\tau}, \ddot{\tau})}$
      \State $c_{2} \gets \prod_{(\dot{s} \sim \ddot{s}) \in s \cap [\dot{\nu}, \ddot{\nu}]} \kappa(v_{r^{\dagger} \cap (\dot{s}, \ddot{s})}^{\dagger} | v_{\braces{\dot{s}, \ddot{s}}}, y_{[\dot{s}, \ddot{s}]}^{\dagger}) \prod_{(\dot{\tau} \sim \ddot{\tau}) \in \tau \cap [\dot{\nu}, \ddot{\nu}]} c_{2}^{(\dot{\tau}, \ddot{\tau})}$
      \State $\varpi_{1} \gets \prod_{(\dot{\tau} \sim \ddot{\tau}) \in \tau^{\dagger} \cap [\dot{\nu}, \ddot{\nu}]} \varpi_{1}^{(\dot{\tau}, \ddot{\tau})}$
      \State $\varpi_{2} \gets \prod_{(\dot{\tau} \sim \ddot{\tau}) \in \tau \cap [\dot{\nu}, \ddot{\nu}]} \varpi_{2}^{(\dot{\tau}, \ddot{\tau})}$
      \State $\alpha_{(\dot{\nu}, \ddot{\nu})}, z_{(\dot{\nu}, \ddot{\nu})}^{*} \gets \Call{2-coin-sep}{c_{1}, c_{2}, \varpi_{1}, \varpi_{2}, z_{(\dot{\nu}, \ddot{\nu})}^{\dagger}, z_{(\dot{\nu}, \ddot{\nu})}}$
      \If{$\alpha_{(\dot{\nu}, \ddot{\nu})}$}
        \State $v_{r^{*} \cap (\dot{\nu}, \ddot{\nu})}^{*}, y_{[\dot{\nu}, \ddot{\nu}]}^{*} \gets v_{r^{\dagger} \cap (\dot{\nu}, \ddot{\nu})}^{\dagger}, y_{[\dot{\nu}, \ddot{\nu}]}
        ^{\dagger}$
      \Else
        \State $v_{r^{*} \cap (\dot{\nu}, \ddot{\nu})}^{*}, y_{[\dot{\nu}, \ddot{\nu}]}^{*} \gets v_{r \cap (\dot{\nu}, \ddot{\nu})}, y_{[\dot{\nu}, \ddot{\nu}]}$
      \EndIf
    \EndFor
    \State $\Call{adapt-hidden-proposal}{\braces{\alpha_{(\dot{\nu}, \ddot{\nu})}: (\dot{\nu} \sim \ddot{\nu}) \in \nu}}$ \Comment{See Section \ref{sec:localize}}
    \State \Return $v_{r^{*}}^{*}, z^{*}, y^{*}$
    \EndFunction
  \end{algorithmic}
\end{breakablealgorithm}

\begin{breakablealgorithm}
  \caption{Random walk Barker-within-Gibbs update for $\pi(\theta | v_{\tau}, z, y)$ (see Section \ref{ssec:param}). \label{alg:param}}
  \begin{algorithmic}
    \Function{update-param}{$\theta, v_{\tau}, z, y$}
    \State $\theta^{\dagger} \sim \kappa(\theta^{\dagger} | \theta)$ \Comment{Typically (tuned) random walk proposal}
    \For{$(\dot{\tau} \sim \ddot{\tau}) \in \tau$}
      \State $c_{1}^{(\dot{\tau}, \ddot{\tau})} \gets \braces{\kappa(\theta | \theta^{\dagger}) \pi(\theta^{\dagger})}^{1/(|\tau| - 1)} h_{\theta^{\dagger}}(y_{\dot{\tau}}, v_{\braces{\dot{\tau}, \ddot{\tau}}})$
      \State $c_{2}^{(\dot{\tau}, \ddot{\tau})} \gets \braces{\kappa(\theta^{\dagger} | \theta) \pi(\theta)}^{1/(|\tau| - 1)} h_{\theta}(y_{\dot{\tau}}, v_{\braces{\dot{\tau}, \ddot{\tau}}})$
      \State $\varpi_{1}^{(\dot{\tau}, \ddot{\tau})} \gets \Call{construct-param-coin}{\theta^{\dagger}, \theta, y_{\dot{\tau}}, v_{\braces{\dot{\tau}, \ddot{\tau}}}}$ \Comment{See Section \ref{ssec:retro}}
      \State $\varpi_{2}^{(\dot{\tau}, \ddot{\tau})} \gets \Call{construct-param-coin}{\theta, \theta^{\dagger}, y_{\dot{\tau}}, v_{\braces{\dot{\tau}, \ddot{\tau}}}}$ \Comment{See Section \ref{ssec:retro}}
    \EndFor
    \State $\alpha, z^{*} \gets \Call{dcbf}{\braces{c_{1}^{(\dot{\tau}, \ddot{\tau})}, c_{2}^{(\dot{\tau}, \ddot{\tau})}, \varpi_{1}^{(\dot{\tau}, \ddot{\tau})}, \varpi_{2}^{(\dot{\tau}, \ddot{\tau})}, z_{(\dot{\tau}, \ddot{\tau})}: (\dot{\tau} \sim \ddot{\tau}) \in \tau}}$
    \If{$\alpha$}
      \State $\theta^{*} \gets \theta^{\dagger}$
    \Else{}
      \State $\theta^{*} \gets \theta$
    \EndIf
    \State $\Call{adapt-param-proposal}{\alpha, \theta, \theta^{\dagger}}$ \Comment{See e.g. \citep{chimisov2018air}}
    \State \Return $\theta^{*}, z^{*}$
    \EndFunction
  \end{algorithmic}
\end{breakablealgorithm}

\begin{breakablealgorithm}
  \caption{2-coin algorithm with separate retrospective sampling (see Section \ref{ssec:retro}). \label{alg:2coinsep}}
  \begin{algorithmic}
    \Function{2-coin-sep}{$c_{1}, c_{2}, \varpi_{1}, \varpi_{2}, z_{1}, z_{2}$}
    \State $z_{1}^{*} \gets z_{1}$
    \State $z_{2}^{*} \gets z_{2}$
    \Loop
      \State $f_{0} \sim \op{Bern}{c_{1} / (c_{1} + c_{2})}$
      \If{$f_{0}$}
        \State $f_{1}, z_{1}^{*} \gets \varpi_{1}(z_{1}^{*})$
        \If{$f_{1}$}
          \Return $1, z_{1}^{*}$
        \EndIf
      \Else
        \State $f_{2}, z_{2}^{*} \gets \varpi_{2}(z_{2}^{*})$
        \If{$f_{2}$}
          \Return $0, z_{2}^{*}$
        \EndIf  
      \EndIf
    \EndLoop
    \EndFunction
  \end{algorithmic}
\end{breakablealgorithm}

\begin{breakablealgorithm}
  \caption{Divide-and-conquer Bernoulli factory (See \citep{stumpf2025scalable}). \label{alg:dcbf}}
  \begin{algorithmic}
    \Function{dcbf}{$\braces{c_{1}^{(i)}, c_{2}^{(i)}, \varpi_{1}^{(i)}, \varpi_{2}^{(i)}, z_{i}}_{i=1}^{n}(, \textsc{depth})$}
    \If{\textsc{depth} = 0}
      \State \Return \Call{2-coin-com}{$\prod_{i} c_{1}^{(i)}, \prod_{i} c_{2}^{(i)}, \prod_{i} \varpi_{1}^{(i)}, \prod_{i} \varpi_{2}^{(i)}, \bigcup_{i} z_{i}$}
    \EndIf
    \State $\braces{z_{i}^{*}}_{i=1}^{n} \gets \braces{z_{i}}_{i=1}^{n}$
    \Loop
      \State $\alpha_{1}, \braces{z_{i}^{*}}_{i=1}^{\floor{n/2}} \gets \Call{dcbf}{\braces{c_{1}^{(i)}, c_{2}^{(i)}, \varpi_{1}^{(i)}, \varpi_{2}^{(i)}, z_{i}^{*}}_{i=1}^{\floor{n/2}}, \textsc{depth} - 1}$
      \State $\alpha_{2}, \braces{z_{i}^{*}}_{i=\ceil{n/2}}^{n} \gets \Call{dcbf}{\braces{c_{1}^{(i)}, c_{2}^{(i)}, \varpi_{1}^{(i)}, \varpi_{2}^{(i)}, z_{i}^{*}}_{i=\ceil{n/2}}^{n}, \textsc{depth} - 1}$
      \If{$\alpha_{1} = \alpha_{2}$}
        \State \Return $\alpha_{1}$
      \EndIf
    \EndLoop
    \EndFunction
  \end{algorithmic}
\end{breakablealgorithm}

\begin{breakablealgorithm}
  \caption{2-coin algorithm with common retrospective sampling (see Section \ref{ssec:retro}). \label{alg:2coincom}}
  \begin{algorithmic}
    \Function{2-coin-com}{$c_{1}, c_{2}, \varpi_{1}, \varpi_{2}$}
    \Loop
      \State $f_{0} \sim \op{Bern}{c_{1} / (c_{1} + c_{2})}$
      \If{$f_{0}$}
        \State $f_{1}, z^{*} \gets \varpi_{1}(z^{*})$
        \If{$f_{1}$}
          \Return $1, z^{*}$
        \EndIf
      \Else
        \State $f_{2}, z^{*} \gets \varpi_{2}(z^{*})$
        \If{$f_{2}$}
          \Return $0, z^{*}$
        \EndIf  
      \EndIf
    \EndLoop
    \EndFunction
  \end{algorithmic}
\end{breakablealgorithm}


\end{document}